\documentclass[10pt]{article}

\usepackage{amsmath}
\usepackage{amssymb}

\usepackage{graphicx}

\usepackage{cite}

\usepackage{color} 

\usepackage[labelfont=bf,labelsep=period,justification=raggedright]{caption}

\makeatletter
\renewcommand{\@biblabel}[1]{\quad#1.}
\makeatother

\date{}

\begin{document}

% Title must be 150 characters or less
\begin{flushleft}
{\Large
\textbf{Chemotactic response and adaptation dynamics\\in {\it Escherichia coli}}
}
% Insert Author names, affiliations and corresponding author email.
\\
Diana Clausznitzer$^{1,2}$, 
Olga Oleksiuk$^{3}$,
Linda L\o{}vdok$^{3}$,
Victor Sourjik$^{3}$,
Robert G.~Endres$^{1, 2\ast}$
\\
\bf{1} Imperial College London, Division of Molecular Biosciences, South Kensington, SW7 2AZ, London, United Kingdom
\\
\bf{2} Centre for Integrated Systems Biology at Imperial College, Imperial College London, London SW7 2AZ, United Kingdom
\\
\bf{3} Zentrum f\"ur Molekulare Biologie der Universit\"at Heidelberg, DKFZ-ZMBH Alliance, Im Neuenheimer Feld 282, 69120 Heidelberg, Germany
\\
$\ast$ E-mail: r.endres@imperial.ac.uk
\end{flushleft}

% Please keep the abstract between 250 and 300 words
\section*{Abstract}

% \subsection*{Background}
Adaptation of the chemotaxis sensory pathway of the bacterium {\it Escherichia coli} is integral for detecting chemicals over a wide range of background concentrations, ultimately allowing cells to swim towards sources of attractant and away from repellents. Its biochemical mechanism based on methylation and demethylation of chemoreceptors has long been known. Despite the importance of adaptation for cell memory and behavior, the dynamics of adaptation are difficult to reconcile with current models of precise adaptation.
% \subsection*{Methodology/Principal Findings}
Here, we follow time courses of signaling in response to concentration step changes of attractant using {\it in vivo} fluorescence resonance energy transfer measurements. Specifically, we use a condensed representation of adaptation time courses for efficient evaluation of different adaptation models. To quantitatively explain the data, we finally develop a dynamic model for signaling and adaptation based on the attractant flow in the experiment, signaling by cooperative receptor complexes, and multiple layers of feedback regulation for adaptation. We experimentally confirm the predicted effects of changing the enzyme-expression level and bypassing the negative feedback for demethylation.
% \subsection*{Conclusion/Significance}
Our data analysis suggests significant imprecision in adaptation for large additions. Furthermore, our model predicts highly regulated, ultrafast adaptation in response to removal of attractant, which may be useful for fast reorientation of the cell and noise reduction in adaptation.

% Please keep the Author Summary between 150 and 200 words
% Use first person. PLoS ONE authors please skip this step. 
% Author Summary not valid for PLoS ONE submissions.   
\section*{Author Summary}

Bacterial chemotaxis is a paradigm for sensory systems, and thus has attracted immense interest from biologists and modelers alike. Using this pathway, cells can sense chemical molecules in their environment, and bias their movement towards nutrients and away from toxins. To avoid over- or understimulation of the signaling pathway, receptors adapt to current external conditions by covalent receptor modification, ultimately allowing cells to chemotax over a wide range of background concentrations. While the robustness and precision in adaptation was previously explained, we quantify the dynamics of adaptation, important for cell memory and behavior, as well as noise filtering in the pathway. Specifically, we study the intracellular signaling response and subsequent adaptation to concentration step changes in attractant chemicals. We combine measurements of signaling in living cells with a dynamic model for strongly coupled receptors, even including the effects of concentration flow in the experiment. Using a novel way of summarizing time-dependent data, we derive a new adaptation model, predicting additional layers of feedback regulation. As a consequence, adaptation to sudden exposure of unfavorable conditions is very fast, which may be useful for a quick reorientation and escape of the cell.

\section*{Introduction}

% 1. importance of adaptation
Cells are able to sense and respond to various external stimuli. To extend the working range of their sensory pathways, biochemical mechanisms allow for adaptation to persistent stimulation, resulting in only a transient response. The dynamics of adaptation are important as they often represent the cells' memory of previous environmental conditions, directly affecting cellular behavior~\cite{JaasJackKea07,MarBibOes95,HillApiSchaf05,ZigSull97,ShiZus94,SpehrHagZuf09,MuzzGomOuden09}. Fast adaptation means that cells stop responding and that their biochemical pathways quickly ``forget'' the stimulus. In contrast, slow adaptation leads to long-lasting effects in the cells. The dynamics of adaptation are often difficult to understand in detail, since they emerge from multiple, simultaneously occurring processes.
In higher organisms, adaptation is best documented in the insect and vertebrate visual system, where multiple mechanisms adjust the receptor sensitivity to ambient light levels. For instance, phototransduction in the vertebrate eye involves up to nine different mechanisms for adaptation~\cite{PughNikLam99}.
However, even in the well-characterized chemotaxis sensory system in {\it Escherichia coli}~\cite{Berg00,FalHaz01,Sou04,WadArm04,BakWolStock06}, adaptation, in particular its molecular mechanism and dynamics, is not well understood. This constitutes a huge deficit as there has recently been immense interest in the chemotactic behavior of bacteria~\cite{ClarkGrant05,VlaLovSou08,EmoClu08,ZonBray09,VladSour09} and noise filtering~\cite{EmoClu08,AndYiIgl06,TuShimBerg08}.
Here, we use adaptation time-course data from {\it in vivo} fluorescence resonance energy transfer (FRET) measurements and  quantitative modeling to address this problem.

% 2. chemotaxis pathway and open questions
The chemotaxis pathway in {\it E. coli} allows cells to sense chemicals and to swim towards more favorable environments by successive periods of straight swimming (running) and random reorientation (tumbling).
Transmembrane chemoreceptors, including the highly abundant Tar and Tsr receptors, cluster at the cell poles and act as ``antennas'' to detect various attractants (e.g. certain amino acids and sugars) and repellents (e.g. certain metal ions) with high sensitivity~\cite{BrayLevMorFir98}.
Receptors activate an intracellular signaling pathway, which results in the phosphorylation of diffusible response regulator CheY (CheY-P) via kinase CheA. CheY-P binds to the flagellated rotary motors and induces tumbling. For details of the pathway see the Supplementary Text S1.
The interactions between different proteins in the chemotaxis pathway during signaling have been well characterized. Specifically, FRET measurements on the response regulator CheY-P and its phosphatase CheZ have elucidated the signaling in the chemotaxis pathway~\cite{SouBerg2002,SouBerg04,KenSou09}.

Adaptation in {\it E. coli} is based on reversible methylation and demethylation of receptors at specific modification sites, catalyzed by enzymes CheR and phosphorylated CheB (CheB-P), respectively. Tar and Tsr receptors have four major modification sites. In addition, the Tsr receptor has two minor modification sites which are methylated less strongly~\cite{ChalWeis05}.
Receptor modification regulates the receptor activity and provides a recording of experienced concentration changes~\cite{Kosh81,LiStock09,VladSour09}.
As a consequence, the rate of tumbling was found to adapt precisely for different ligand concentrations~\cite{BergBrown72,AlonSurLei99}. To achieve the return of the receptor activity to its pre-stimulus value, receptor activity-dependent phosphorylation of CheB provides a negative feedback on the receptor activity. In addition, the rates of methylation and demethylation depend on the receptor activity~\cite{ToeGoyStock79,LiHaz06,LaiBarBar06}, representing further layers of feedback regulation.
To modify receptors, CheR and CheB molecules can bind to a specific tether sequence at the carboxyl-terminus of Tar and Tsr receptors, and act on several nearby receptors, so-called assistance neighborhoods~\cite{LiHaz05}. This is believed to compensate for the low numbers of CheR and CheB (hundreds of molecules)~\cite{LiHaz04}, although larger numbers have been reported~\cite{SimStockStock87}.

Although a lot is known about the components of the chemotaxis pathway, several open questions remain to be answered in adaptation. ({\it i}) Despite their importance for cell behavior, memory and noise filtering, the dynamics of adaptation and the methylation level are largely unknown. This is because the methylation level is difficult to measure precisely, relying on quantification of receptor protein and radioactively-labeled methylation substrate (methionine) incorporated into receptors~\cite{KortGoyAdl75,ChelGutKosh84,LaiHaz05,ChalWeis05}.
So far, only the initial rate of adaptation was inferred from the rate of change in motor bias in response to saturating amounts of added attractant~\cite{BergBrown72}. ({\it ii}) The molecular mechanism of adaptation, in particular demethylation, remains unclear.
While CheR binds strongly to the tether, suggested to increase its concentration in the vicinity of methyl-accepting sites~\cite{WuLiWeis96}, the binding affinity of \mbox{CheB} was found to be very low~\cite{BarBarHaz02}.
Instead, binding of CheB-P to the tether induces an allosteric activation of the receptor, increasing the demethylation rate~\cite{BarBarHaz02}.
Furthermore, while the receptor activity-dependence of the methylation and demethylation rates is believed to be a requirement for robust precise adaptation (see below), it is not known if adaptation is precise at the receptor level.
Time-course data from {\it in vivo} FRET experiments, monitoring receptor activity upon stimulation, is ideally suited to study the adaptation dynamics and address these questions.

% modelling results
Extensive mathematical modeling has described different aspects of the chemotaxis pathway. However, modeling has mainly focused on explaining the initial response to addition of attractant, as well as precise adaptation, i.e. the complete return of the signaling activity to pre-stimulus level long after the stimulus.
Briefly, the Monod-Wyman-Changeux (MWC) model was used to successfully describe the signaling of two-state receptor complexes, formed by \mbox{10-20}~strongly interacting receptor dimers~\cite{SouBerg04,MelTu05,KeyEndSko06,EndWin06,EndSouWin2008}. In this model, receptor-receptor coupling provides a mechanism for signal amplification and integration. Alternative receptor models are outlined in the {\it Discussion}.
Furthermore, Barkai and Leibler showed that precise adaptation is robust (insensitive to parameter variations in the pathway), if the kinetics of receptor methylation depends only on the activity of receptors and not explicitly on the receptor methylation level or external chemical concentration~\cite{BarLei97}. 
Their idea was later extended by others, providing conditions for precision~\cite{YiHuaDoy00, MelTu03}, as well as robustness to noise by the network architecture~\cite{KolLovSou05} and assistance neighborhoods~\cite{EndWin06,HanEndWin08}. Most recently, adaptation to exponential ramps and sinusoidal concentration changes was investigated~\cite{TuShimBerg08}.
However, none of these studies have directly compared to adaptation time-courses from FRET.

% summary of paper, Fig 1
Here, we use {\it in vivo} FRET data obtained from cells adapted to ambient concentrations of attractant $\alpha$-methylaspartate (MeAsp; a non-metabolizable variant of amino acid aspartate) and stimulated in a flow chamber by various concentration step changes~\cite{SouBerg2002}. Recording the average initial response amplitudes for each added and, after adaptation, removed concentration step change results in dose-response curves (Fig.~\ref{fig:DR_MWC_and_adaptation}, symbols).
We use a dynamic version of the MWC model, which, in addition to mixed complexes of Tar and Tsr receptors, 
includes a more detailed description of the adaptation dynamics than used in previous models of chemotaxis. Specifically, we predict multiple layers of feedback regulation during adaptation, especially for demethylation by CheB. In addition, we take into account the kinetics of attractant flow in FRET experiments. 
This allows us to quantitatively describe dose-response curves (Fig.~\ref{fig:DR_MWC_and_adaptation}, lines), in particular the observed reduced response amplitudes for removal of MeAsp, which previously could not be explained by the MWC model ({\it Inset}).
To analyze the adaptation dynamics, we use the data collapse, a condensed representation of time courses. This data collapse enables us to evaluate the effect of ligand flow and adaptation imprecision, to infer the kinetics of the receptor methylation level, as well as to efficiently compare adaptation models from the literature to experimental data. Finally, we experimentally test two predictions to validate our adaptation model.
We change the adapted receptor activity, and use a non-regulatable CheB mutant to bypass its negative feedback on the receptor activity. Our combined study of experiments and modeling shows that adaptation is relatively imprecise at the receptor level for large stimuli, and that demethylation is more tightly regulated than previously thought. This leads to very short tumbles for sudden occurrences of unfavorable conditions, allowing cells to quickly reorient their swimming direction after a short tumble.

% Results and Discussion can be combined.
\section*{Results}

\subsection*{Dynamic MWC model for {\it in vivo} FRET response\mbox{ }}

Our dynamic MWC model, described in the following, combines the previously used MWC model for receptor signaling by strongly-coupled receptor complexes (denoted here by static MWC model), with the dynamic effects of adaptation by receptor modification, as well as ligand concentration flow.
In the static MWC model, mixed receptor complexes composed of Tar (aspartate receptor) and Tsr (serine receptor, which also binds aspartate with lower affinity) are considered in their {\it in vivo} ratio.
Using a two-state assumption, the activity of a receptor complex is given by its probability to be in {\it on} (active), which depends on the free-energy difference $F=F_{\text{on}}-F_{\text{off}}$ between its {\it on} and {\it off} (inactive) state~\cite{KeyEndSko06,EndSouWin2008},
\begin{equation}
 A \equiv p_{\text{on}} = \frac{e^{-F_\text{on}}}{e^{-F_\text{on}} + e^{-F_\text{off}}} = \frac{1}{1+e^{F}}.
\end{equation}
This free-energy difference, $F(m, c)$, is determined by two contributions, one from methylation (in terms of receptor methylation level $m$) favoring the {\it on} state, and one from attractant binding at MeAsp concentration $c$ favoring the {\it off} state.
The free-energy difference $F$ also depends on several parameters such as free-energy difference per added methyl group, the number $N$ of receptor dimers in a complex, as well as the ligand dissociation constants $K_{a (s)}^{\text{on}}$ and $K_{a (s)}^{\text{off}}$ for Tar (Tsr) receptors in their {\it on} and {\it off} states, respectively.
Most of these parameters were determined previously (see {\it Materials and Methods}). Similar free-energy based two-state models were recently used to describe clustering of ion channels~\cite{UrsHuaPhil07} and small GTPases in eukaryotic cells~\cite{GurKahEnd09}. In the new dynamic MWC model, we include the effects of variable receptor complex sizes, adaptation dynamics, and MeAsp concentration flow on the initial response to concentration changes.

\begin{figure*}[!ht]
  \centering
     \includegraphics[]{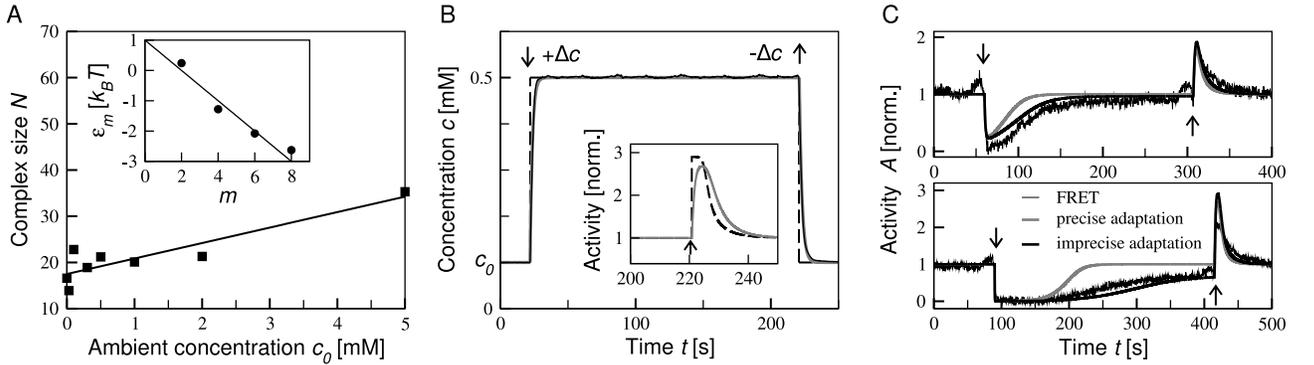}
\caption{\label{fig:methods}{\bf Model ingredients.}
(A)~Size of adapted receptor complex $N$ (total number of Tar and Tsr receptors per complex) as function of ambient concentration $c_0$ (corresponding to adapted methylation level $m$). Individual complex sizes (symbols) were obtained by fitting MWC model to dose-response curves for addition of MeAsp. These values were fitted by a linear function (line).
(A~{\it Inset})~Energy contribution to receptor complex free energy from methylation level $m$ per receptor dimer. Shown are fitting parameters for Tar receptors from~\cite{EndSouWin2008} (symbols), as well as linear fit $\epsilon(m)\!\!=\!\!1-0.5 m$ (in units of $k_B T$ with $k_B$ the Boltzmann constant and $T$ absolute temperature).
(B)~Profile of concentration step change as measured experimentally using fluorescent marker (solid black line)~\cite{SouBerg2002}, exponential fit used in dynamic MWC model for WT1 MeAsp profile (gray line; rate constants $\lambda_{\text{add}}\!\!=\!\!0.6/s$ and $\lambda_{\text{rem}}\!\!=\!\!0.5/s$), and perfect step change (black dashed line). Addition and removal times are marked by arrows.
(B {\it Inset})~Response of mixed receptor complex to MeAsp removal for perfect (black dashed line) and exponentially fitted step change (gray line).
(C)~Typical time courses of receptor complex activity in response to two different MeAsp concentration step changes, $\Delta c\!\!=\!\!0.05\,\text{mM}$ (top) and $\Delta c\!\!=\!\!0.4\,\text{mM}$ (bottom), at ambient concentration $c_0\!\!=\!\!0.1\,\text{mM}$. Experimental FRET measurement (thin black line), as well as dynamic MWC model for precise (gray lines) and imprecise adaptation (black lines; $m_{max}\!\!=\!\!4.1$ and $K\!\!=\!\!0.5$).
FRET and receptor complex activities were normalized by adapted pre-stimulus values before addition of MeAsp.
}
\end{figure*}

The dependence of the receptor complex size on the ambient concentration and hence methylation level was determined as follows:
First, the receptor complex size was obtained for each ambient concentration using a least-squares fit to addition dose-response curves (see Fig.~\ref{fig:methods}A and {\it Materials and Methods}). Consistent with previous modeling results, we find that the receptor complex size increases with increasing ambient concentration~\cite{MelTu07,EndSouWin2008}. As the simplest assumption, we used a linear relationship between receptor complex size and ambient concentration (Fig.~\ref{fig:methods}A), allowing us to interpolate the receptor complex size for removal dose-response curves.
Analyzing the signaling pathway of {\it E. coli}, we also found the phosphorylation reactions are sufficiently fast to assume that concentrations of phosphorylated (and unphosphorylated) proteins are in quasi-steady state. Furthermore, the concentrations of activated proteins are approximately proportional to the receptor complex activity. Both these conditions allow us to use the receptor complex activity as a substitute for the down-stream activity measured by FRET reducing the number of model parameters for fitting to data greatly~(see Supplementary Text S1). This approximation was also used in previous work, but was never explicitly tested~\cite{EndWin06,KeyEndSko06,EndSouWin2008}.

Adaptation occurs on a similar time scale as the kinetics of the MeAsp concentration flow. In experiments, changes in MeAsp concentration are established over several seconds, due to the finite flow speed and mixing effects in the flow chamber.
In our model, we assume exponentially rising and falling concentration changes upon addition and removal in line with previous measurements by Sourjik and Berg~(Fig.~\ref{fig:methods}B)~\cite{SouBerg2002}.
Adaptation is mediated by methylation and demethylation enzymes CheR and CheB, respectively. The process is described by the kinetics of the average receptor methylation level $m$ in a receptor complex,
\begin{equation}
 \frac{dm}{dt}=g_R (1-A) - g_B A^3\label{eq:preciseadaptation},
\end{equation}
where the adapted receptor-complex activity $A^*$ is determined by the steady-state condition $dm/dt = 0 = g_R (1-A^*) - g_B {A^*}^3$. According to our model, receptors are methylated when the complex is inactive, and demethylated when it is active. Furthermore, we postulate a very strong sensitivity of the demethylation rate on activity, possibly due to cooperativity of CheB-P molecules. This mechanism explains the strong asymmetry, which is observed in experimentally measured time courses (cf.~Fig.~\ref{fig:methods}C) where adaptation of inactive receptors (methylation) is slow compared to the rapid adaptation of active receptors (demethylation).
Hence, during a concentration step change the initial response amplitude of receptor complexes is reduced by simultaneous adaptation, which is particularly important for removal of concentration (see Fig.~\ref{fig:methods}B~{\it Inset}). Note that the asymmetry between slow adaptation of inactive and active receptors, respectively, cannot simply be changed by adjusting the rate constants of methylation and demethylation individually, since they are constrained by the adapted activity $A^*$.
For details of this adaptation model see {\it Materials and Methods}, and for a potential molecular mechanism of demethylation, see {\it Discussion}.

\begin{figure}[!ht]
  \centering
     \includegraphics[]{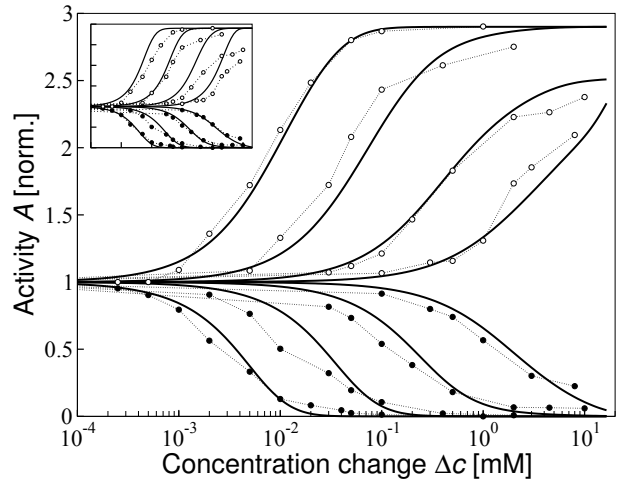}
\caption{\label{fig:DR_MWC_and_adaptation}{\bf Response of wild-type cells to step changes $\Delta c$ of MeAsp concentration at different ambient concentrations.}
Dose-response curves: Symbols represent averaged response from FRET data (WT1) after adaptation to ambient concentrations 0, 0.1, 0.5 and 2~mM as  measured by Sourjik and Berg~\cite{SouBerg2002} (filled and open circles correspond to response to addition and removal of attractant, respectively). Solid lines represent the dynamic MWC model of mixed Tar/Tsr-receptor complexes including ligand rise (addition) and fall (removal), as well as adaptation (receptor methylation) dynamics.
({\it Inset})~Dose-response curves for MWC model without adaptation dynamics (lines).
FRET and receptor complex activities were normalized by adapted pre-stimulus values at each ambient concentration.
Squared errors between model and experimental data for the four dose-response curves shown are 0.64 (dynamic MWC model) and 3.95 (static MWC model), respectively. For comparison, fitting to eight addition and removal dose-response curves using $K_{a (s)}^{\text{on}}$, $K_{a (s)}^{\text{off}}$, as well as a linear function $N(c_0)$ as fitting parameters, yields squared errors 0.83 (dynamic MWC model) and 2.09 (static MWC model), see Supplementary Text S1.
}
\end{figure}

% Fig. 1
Experimental dose-response curves (Fig.~\ref{fig:DR_MWC_and_adaptation}, symbols) describe the initial response of adapted wild-type cells to sudden changes (addition and removal) in MeAsp concentration~\cite{SouBerg2002}. These responses are taken from time courses measured by {\it in vivo} FRET (cf. Fig.~\ref{fig:methods}). Additional, previously unpublished dose-response curves are provided in the Supplementary Text S1. For details of the experiments see {\it Material and Methods}.
Our dynamic MWC model, which includes the effects of adaptation and MeAsp flow, quantitatively describes the experimental dose-response curves.
Specifically, adaptation leads to a non-saturated response for large MeAsp step changes $\Delta c$ at high ambient concentrations, which is not seen in the static MWC model without adaptation dynamics~(Fig.~\ref{fig:DR_MWC_and_adaptation}~{\it Inset}). Note, however, that responses eventually do saturate according to the dynamic MWC model for even larger concentration step changes due to the presence of Tsr receptors (at 0.5 mM ambient for $\Delta c\!>\!40\, \text{mM}$; not shown).  
The dynamic MWC model describes the dose-response data significantly better than the static MWC model, as indicated by their overall squared errors in the caption of Fig.~\ref{fig:DR_MWC_and_adaptation}, as well as residual errors detailed in the Supplementary Text S1.
The receptor-complex activity, as well as FRET data were normalized by their adapted pre-stimulus values at ambient concentration to compare model and experimental data (see {\it Materials and Methods}).

\subsection*{Data collapse of time courses for adaptation dynamics}%

\begin{figure*}[!ht]
  \centering
     \includegraphics[]{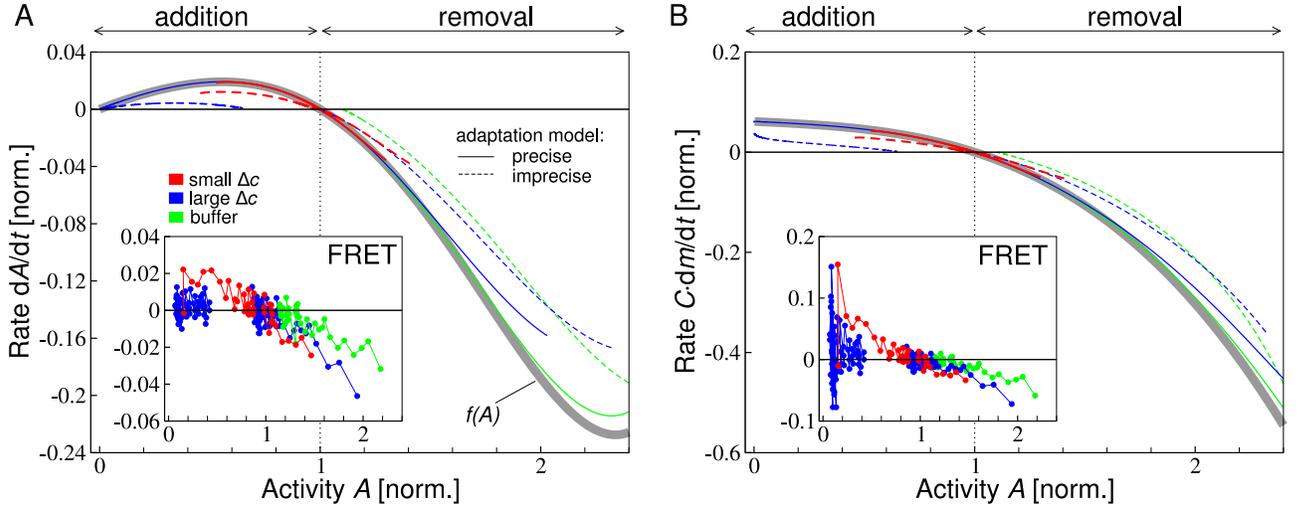}
\caption{\label{fig:dAdm}{\bf Adaptation dynamics as function of receptor activity for WT1 at ambient concentration $c_0\!\!=\!\!0.1\,\text{mM}$ for addition and subsequent removal of small (red lines and symbols) and large (blue lines and symbols) MeAsp concentration step changes, as well as removal of MeAsp back to zero ambient concentration (buffer; green lines and symbols).}
(A)~Rate of change of receptor complex activity $dA/dt$ as occurs during adaptation. Thick gray line is the analytical result from the dynamic MWC model, where activity change is purely from adaptation (methylation) $dA/dt\!\!=\!\!(dA/dm) (dm/dt)\!\!=\!\!f(A)$. Colored lines show results from simulated time series for small ($\Delta c\!\!=\!\!0.03\,\text{mM}$) and large ($\Delta c\!\!=\!\!0.4\,\text{mM}$) concentration step changes in MeAsp concentration, with activity dynamics recorded starting 10 s after the onset of concentration step change. Precise (solid lines), as well as imprecise adaptation (dashed lines; $m_{max}\!\!=\!\!4.1$ and $K\!\!=\!\!0.5$) are considered.
(A~{\it Inset})~Rate of FRET activity change from experimental time-course data. Small ($\Delta c\!\!=\!\!0.03\,\text{mM}$) and large ($\Delta c\!\!=\!\!2\,\text{mM}$) concentration step changes.
(B)~Rate of change of the methylation level $dm/dt$ corresponding to panel A (normalized by adapted activity $A^*$ and $C\!\!=\!\! N/2 $, where $N$ is the receptor complex size).
Effective rate of change of methylation level for all time courses is obtained by $(dA/dt) / [A (1-A)]$.
(B~{\it Inset})~Effective rate of change of methylation level from experimental time-course data. 
FRET and receptor complex activities, as well as activity rate changes were normalized by adapted pre-stimulus activities at ambient concentrations before addition of MeAsp.
}
\end{figure*}

The short-term behavior in the time-course data (Fig.~\ref{fig:methods}C) is essential in deriving our adaptation model, used to quantitatively describe dose-response curves (Fig.~\ref{fig:DR_MWC_and_adaptation}).
Can our adaptation model also describe the long-term behavior in the time-course data, and hence the complete adaptation dynamics?
Our model for precise adaptation predicts that the observable rate of activity change is given by 
\begin{equation}
 \frac{dA}{dt} = \frac{\partial A}{\partial m}\frac{dm}{dt} + \frac{\partial A}{\partial c}\frac{dc}{dt},\label{eq:dA}
\end{equation}
where the rate of change of the methylation level $dm/dt$ is described by Eq.~\ref{eq:preciseadaptation}, and $dc/dt$ is the rate of change of the MeAsp concentration. After a concentration step change, the MeAsp concentration is constant with $dc/dt\!\! =\!\! 0$, and the rate of activity change is given by
\begin{equation}
  \frac{dA}{dt} = \frac{\partial A }{ \partial m } \frac{ dm }{ dt } = A (1-A) \frac{N}{2} \left[ g_R (1-A) - g_B A^3 \right] \equiv f(A),
\end{equation}
where we used that $\partial A / \partial m \!\! = \!\! (\partial A / \partial F)(\partial F / \partial m) \!\! = \!\! A (1-A) N/2$ (see {\it Material and Methods}). Hence, the rate of activity change is a function $f(A)$ of the activity only, independent of ligand concentration and receptor methylation level (except for the minor dependence of the receptor complex size on the ligand concentration, see Supplementary Text S1).
This predicts a data collapse of all adaptation time courses, independent of the duration, size and number of concentration step changes, onto a single curve $dA/dt\!\!=\!\!f(A)$ (Fig.~\ref{fig:dAdm}A, thick gray line).
This non-monotonous function of the activity has three fixed points: the adapted activity $A\!\!=\!\!A^*$, where methylation and demethylation rates exactly balance each other, as well as $A\!\!=\!\!0$ and $A\!\!=\!\!1$, where the receptor complex activity is saturated in the {\it off} and {\it on} state, respectively.
Figure~\ref{fig:dAdm}A~{\it Inset} shows the experimental rate of activity change as extracted from our quantitative time-course data from FRET for different concentration step changes at an ambient concentration.
We observe that, in contrast to the prediction of the model, the rate of activity change depends on the magnitude of the concentration step changes. For addition of large concentration step changes (blue symbols), the rate is reduced and the activity stays below the pre-stimulus value. Furthermore, for total removal of MeAsp concentration (replacement with buffer medium, green symbols), the magnitude of the rate is reduced and the activity remains above the pre-stimulus value.

To explain the deviations from the predicted data collapse, we consider the effects of MeAsp flow and imprecise adaptation in our model. According to Eq.~\ref{eq:dA}, each of the two effects contribute independently to the rate of activity change. 
First, we include the MeAsp flow for concentration step changes as described, and simulate time courses based on the precise adaptation model (Fig.~\ref{fig:dAdm}A, solid lines). We find that in the demethylation regime (negative rate of activity change), the kinetics of concentration step removal gives rise to minor deviations from the curve $f(A)$ in qualitative agreement with experiment. However, in the methylation regime (positive rate of activity change), unlike the experimental data, all time courses lie accurately on the $f(A)$ curve.
Next, we consider imprecise adaptation, i.e. the incomplete return of the activity to pre-stimulus level, which is apparent in the time courses (Fig.~\ref{fig:methods}C and Supplementary Text S1 for quantification). In our model for imprecise adaptation, Eq.~\ref{eq:impreciseadaptation} in {\it Materials and Methods}, the kinetics of the methylation level $dm/dt$ depends explicitly on the receptor methylation level, which leads to significant deviations from the data collapse (Fig.~\ref{fig:dAdm}A, dashed lines). Adaptation after addition of increasing concentration step changes results in a reduced adapted receptor complex activity~(adapted activity after removal is always the same as the concentration is the ambient concentration). Total removal of MeAsp concentration (buffer) results in an increased adapted activity. 
Our imprecise adaptation model is in line with the experimental data, showing that the data collapse is an effective way to compare experimental and theoretical time courses and to quantify the effects of ligand flow and imprecise adaptation. 
We also studied the effect of changes in receptor-complex size on the data collapse, which we found to be minor for the concentrations considered here (see Supplementary Text S1).% Fig.~8). }

In addition to the adaptation dynamics, the data collapse allows us to determine the kinetics of the receptor methylation level, which is difficult to measure directly. Figure~\ref{fig:dAdm}B shows the rate of change of the methylation level as a function of the receptor complex activity for experimental data, as well as the dynamic MWC model.
The data and curves were obtained by dividing the rate of activity change $dA/dt$ following concentration step changes by $A (1-A)$. If the activity change is caused only by the adaptation dynamics, this procedure yields a function proportional to the rate of change of the methylation level, $dm/dt$.
According to our precise adaptation model Eq.~\ref{eq:preciseadaptation}, the rate of change of the methylation level is a monotonically decreasing function of activity with one steady state, marking the adapted receptor complex activity (Fig.~\ref{fig:dAdm}B, thick gray line).
Corresponding to the rate of activity change in Fig.~\ref{fig:dAdm}A, the kinetics of ligand flow upon concentration step changes, as well as imprecise adaptation result in deviations from this curve.
As before, we mainly find signatures of imprecise adaptation in the experimental data in Fig.~\ref{fig:dAdm}B~{\it Inset}.

\subsection*{Comparison of different adaptation models}%
\begin{figure*}[!ht]
  \centering
     \includegraphics[]{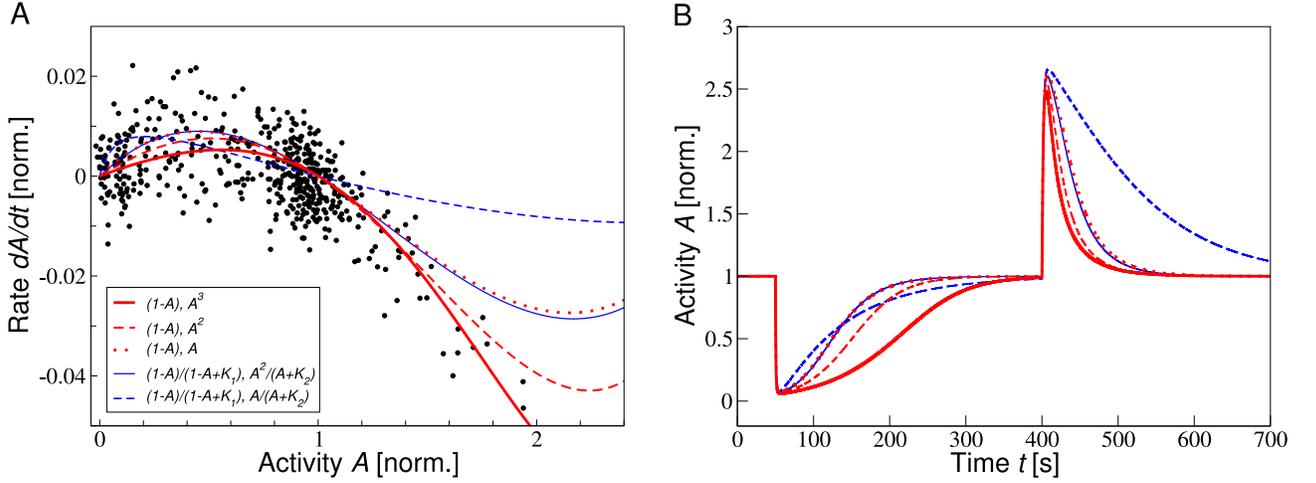}
\caption{\label{fig:compareadaptationmodels}{\bf Comparison of different adaptation models.}
(A)~Rate of activity change during adaptation as a function of activity for FRET data (WT1; symbols) and different adaptation models (colored lines). Experimental FRET activity change is measured at ambient concentration $c_0\!\!=\!\!0.1\,\text{mM}$ for added and subsequently removed concentration step changes $\Delta c$=0.03, 0.05, 0.1, 0.4 and 2 mM. 
For the five models, the dependencies of the methylation and demethylation rates on the receptor complex activity $A$ are given in the legend and are explained in the text.
Models were fitted to the experimental $dA/dt$ data using a least-squares fit, where the methylation rate constant $g_R$ was the only fitting parameter. The demethylation rate $g_B$ was determined to produce the adapted activity $A^*\!\!\approx\!\!1/3$. The parameters $K_1$ and $K_2$ were converted from~\cite{EmoClu08}.
(B)~Representative time courses for the different models in panel~A for a concentration step change $\Delta c\!\!=\!\!0.1\,\text{mM}$ at ambient concentration $c_0\!\!=\!\!0.1\,\text{mM}$.
FRET and receptor complex activities, as well as activity rate changes were normalized by adapted pre-stimulus activities at ambient concentrations before addition of MeAsp.
Least-squares errors between experimental data and model in panel A are 0.0021 [$1-A, A^3$], 0.0022 [$1-A, A^2$], 0.0025 [$1-A, A$], 0.0025 [$(1-A)/(1-A+K_1), A^2/(A+K_2)$], and 0.0036 [$(1-A)/(1-A+K_1), A/(A+K_2)$].
}
\end{figure*}

The data collapse of experimental time courses enables the efficient evaluation of different adaptation models, including our model and other models from the literature (Fig.~\ref{fig:compareadaptationmodels}A).
All models considered here show precise adaptation with the rates of methylation and demethylation only depending on the receptor complex activity, and the explicit activity dependencies given respectively by the first and second term in the legend of Fig.~\ref{fig:compareadaptationmodels}. For instance, the first model (solid red line) is given by Eq.~\ref{eq:preciseadaptation}.
% Common to all models considered here is that the rates of methylation and demethylation only depend on the receptor complex activity, with the explicit activity dependencies given respectively by the first and second term in the legend of Fig.~\ref{fig:compareadaptationmodels}. For instance, the first model (solid red line) is given by Eq.~\ref{eq:preciseadaptation}.
%
Two classes of models are analyzed here. The first class of models, including our model, is based on a linear activity-dependence of the methylation and demethylation rates for binding of CheR and CheB to inactive and active receptor, respectively. Feedback from the activity-dependent phosphorylation of CheB is accounted for by additional factors of the receptor complex activity. 
Our model includes cooperative CheB feedback (solid red line), while linear CheB feedback (dashed red line) and no CheB feedback (dotted red line) are considered as well~\cite{EndWin06,HanEndWin08,VlaLovSou08,KalTuWu09}.
Another class of models has been proposed, showing ultrasensitivity with respect to CheR and CheB protein levels. In these models, the activity-dependence of the methylation and demethylation rates for enzyme binding is described by Michaelis-Menten kinetics, and linear CheB feedback (solid blue line) and no CheB feedback (dashed blue line) is considered~\cite{EmoClu08}. The last model has the property that the rate of change of methylation level becomes independent of activity around the steady-state, leading to extremely long adaptation times.
Details of the alternative adaptation models and the fitting procedure are given in the Supplementary Text S1.
While several models are consistent with the experimental data, our model compares most favorably. 
The ultrasensitive Michaelis-Menten model without CheB feedback seems not to be consistent with the data. 
Comparing simulated time courses for the different adaptation models in Fig.~\ref{fig:compareadaptationmodels}B, our model is best to capture the experimentally observed asymmetry between adaptation to addition and removal of concentration step changes.
The quality of fit between the respective models and data is indicated by their least-squares errors in the caption of Fig.~\ref{fig:compareadaptationmodels}.

\subsection*{Predictions}%
\begin{figure*}[!ht]
  \centering
     \includegraphics[]{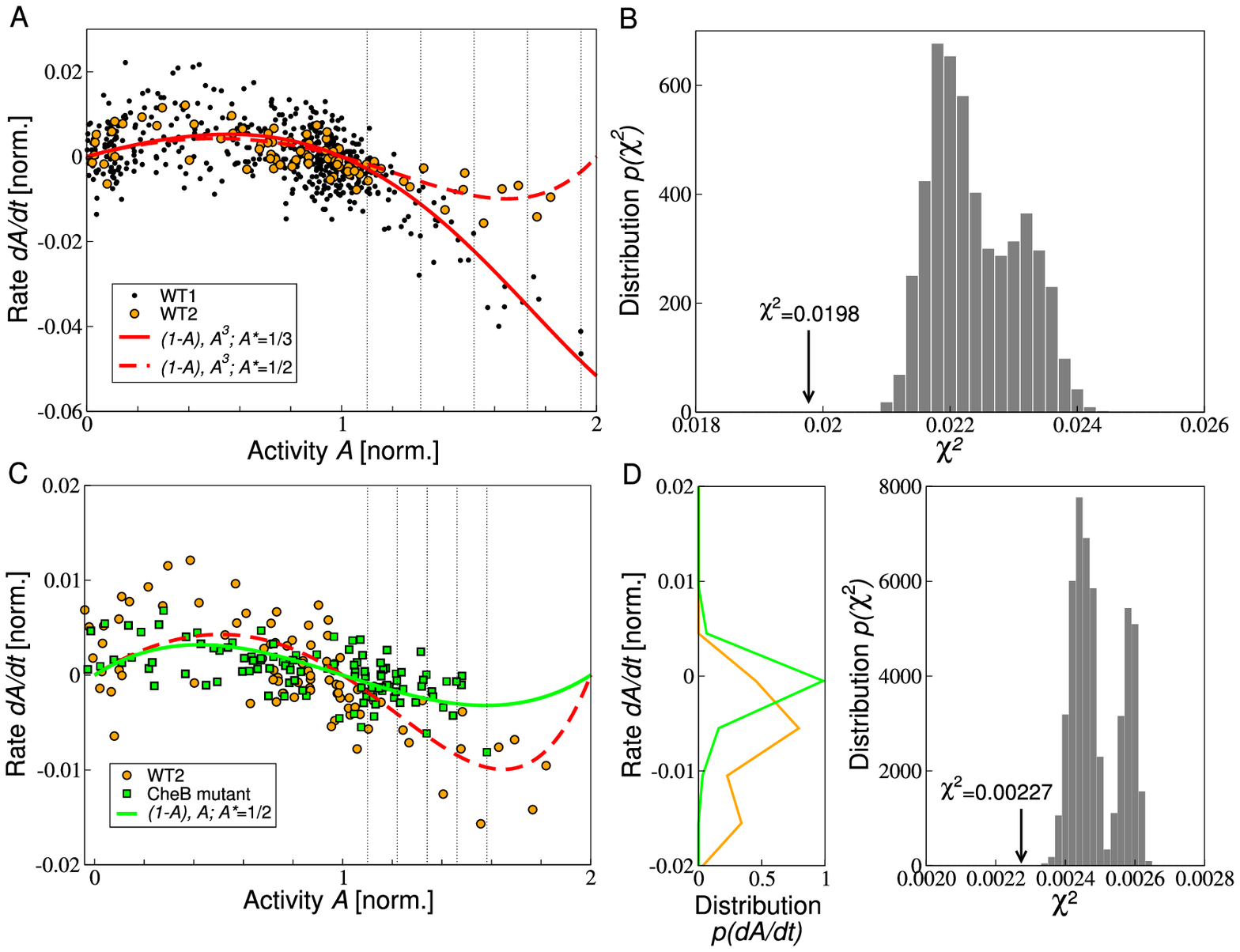}
\caption{\label{fig:dA_flowCheB_effects}{\bf Effects of (A) steady-state activity and (C) CheB regulation by phosphorylation.}
(A) Black and orange dots correspond to the rate of FRET activity change from experimental time-course data for WT1 (Fig.~\ref{fig:compareadaptationmodels}) and for WT2 (addition and subsequent removal of concentration step change $\Delta c\!\!=\!\!0.03\,\text{mM}$ at zero ambient concentration), respectively.
Red lines correspond to the predicted rate of activity change $dA/dt \!\!=\!\! f(A)$ purely from adaptation (solid and dashed lines correspond to steady-state activities $A^*\!\approx\! 1/3$ and $1/2$, respectively). The methylation rate constant $g_R\!\!=\!\!0.0019\,\text{s}^{-1}$ is the same in each case.
Dotted lines indicate bins used to quantify the difference between data sets in panel B. 
(B) Distribution of squared errors $(\chi^2)$ between predicted rate of activity change and experimental data sets for WT1 and WT2, when randomly permuting $10^4$ pairs of data points between the data sets, one pair chosen within each bin in panel A. The error is calculated as the sum of errors for each data set (including the permuted data points) against its respective model. The error of the unpermuted data sets is indicated by the arrow.
(C) Green squares represent the rate of FRET activity change from experimental time-course data for CheB mutant (addition and subsequent removal of concentration step changes $\Delta c\!\!=\!\!0.03\,\text{mM}$ and 0.1~mM at zero ambient concentration). The green line represents the rate of change of receptor complex activity purely from adaptation. Orange dots and red dashed line are the same as in panel A.
Dotted lines indicate bins used to quantify the difference between data sets in panel D. 
(D, left) Distribution of data points of the rate of activity change for activities above $A\!\!=\!\!1.1$ WT2 and CheB mutant data in panel C. 
(D, right) Distribution of squared errors between predicted rate of activity change and experimental data sets for WT2 and CheB mutant, when randomly permuting $10^4$ pairs of data points between the data sets, one pair chosen within each bin in panel C. The error is calculated as the sum of errors for each data set (including the permuted data points) against its respective model. The error of the unpermuted data sets is indicated by the arrow.
}
\end{figure*}

To further validate our adaptation model, we experimentally tested two predictions.
First, in our precise-adaptation model the data collapse depends strongly on the steady-state activity. For instance, increasing the steady-state activity from $A^*\!\!\approx\!\!1/3$ to 1/2 changes the data collapse from the solid to the dashed red line in Fig.~\ref{fig:dA_flowCheB_effects}A. Such an increase in the steady-state activity can be achieved by decreasing CheB expression level, corresponding to a decreasing demethylation rate, at constant CheR expression level.
To validate this prediction, a different wild-type strain (WT2) was created, in which CheB expression was induced from a plasmid, while all other chemotaxis proteins were expressed as before (WT1). The steady-state activity was estimated to be $A^*\!\!\approx\!\!1/2$ (compared to 1/3 in WT1). For details of the strains, see {\it Materials and Methods}.
The data collapse (Fig.~\ref{fig:dA_flowCheB_effects}A, orange circles) corresponds well to the predicted curve (dashed red line).
Second, the activity-dependence of the demethylation rate is diminished according to Eq.~\ref{eq:CheBadaptation} when considering adaptation without feedback through activity-dependent CheB phosphorylation, while keeping the steady-state activity constant (Fig.~\ref{fig:dA_flowCheB_effects}C, green line).
To validate this prediction, a mutant strain was created, which contains non-regulatable CheB with about 10 percent of CheB-P activity. The CheB expression level was increased to produce the kinase activity of WT2 ($A^*\!\!\approx\!\!1/2$). All other chemotaxis proteins are expressed as in WT2 cells.
We find that the experimental rate of FRET-activity change from time-course data (green squares) is consistent with this prediction.

The statistical significance for each of the two predictions (Fig.~\ref{fig:dA_flowCheB_effects}A and C) was tested as follows: For each prediction, we randomly permuted a number of data points from the control experiment and the prediction-testing experiment. Then we calculated the distribution of squared errors between the rates of activity change from the model and FRET measurement for the permuted data sets (Fig.~\ref{fig:dA_flowCheB_effects} B and D). For four permuted pairs of data points the error is always above the error for the unpermuted data sets (Fig.~\ref{fig:dA_flowCheB_effects}). For fewer permutations the error lies at the lower bound of the distribution (not shown). This confirms that the control and prediction-testing data sets are significantly different and match our model.

\section*{Discussion}

%general sentence
Sensing and adaptation are fundamental biological processes, enabling cells to respond and adjust to their external environment. Adaptation extends the range of stimuli a sensory pathway can respond to, while its dynamics determines how long a stimulus will affect the cell's behavior.
%
%1. summary
In this work, we developed a model to quantitatively describe experimental dose-response curves, as well as time courses of chemotaxis signaling in adapting wild-type cells.
Our model includes~({\it i})~the signaling activity of two-state mixed chemoreceptor complexes in response to added or removed attractant concentration step changes based on the Monod-Wyman-Changeux model, ({\it ii}) the kinetics of the ligand concentration in the flow chamber, and ({\it iii}) a detailed mechanism for adaptation, including multiple layers of feedback regulation and imprecise adaptation.
In particular, we find that the finite ligand flow speed and fast, activated demethylation explains for the first time the gradually reduced amplitudes in removal dose-response curves for increasing ambient concentrations (Fig.~\ref{fig:DR_MWC_and_adaptation}).
Our adaptation model introduces a strong receptor-activity dependence of the demethylation rate, and hence is able to reproduce the observed asymmetry of slow adaptation to addition of attractant and fast adaptation to removal of attractant (Fig.~\ref{fig:methods}C). Such dynamics yields long runs up the gradient and short tumbles, sufficient for random reorientation of the cell and escape from potential toxins.
Furthermore, this strong activity dependence has the additional benefit of reducing the fluctuations in receptor methylation level introduced by the adaptation mechanism itself. We found for the total variance of the receptor-complex methylation level $\langle \delta M^2 \rangle\!\!=\!\!0.87$ compared to 2 for a previous model for precise adaptation with weaker activity dependence (details of the calculation can be found in the Supplementary Text S1).
This is because a fluctuation in the receptor methylation level leads to an increased change in activity and hence increased rate to return to the adapted activity.

% data collapse
Our model for precise adaptation predicts the data collapse of adaptation time-courses, allowing the convenient study of the adaptation dynamics (Fig.~\ref{fig:dAdm}A). Specifically, the data collapse allows to evaluate the effects of ligand flow and adaptation dynamics, as well as imprecise adaptation. We found that adaptation to large concentration step changes is significantly imprecise (see Supplementary Text S1 for further details).
We also extracted the kinetics of the receptor methylation level $dm/dt$ from experimental time courses from the data collapse (Fig.~\ref{fig:dAdm}B), which is difficult to measure directly when relying on the quantification of the receptor methylation level using standard biochemical methods~\cite{LaiHaz05,ChalWeis05}. According to our model, the activity-dependence of the receptor methylation level is a monotonously decreasing function of the receptor complex activity.
Ultimately, this kinetics determines the compromise between long memory of previous concentrations and quick recovery for sensing new concentration changes~\cite{ClarkGrant05}.
Furthermore, we experimentally tested two predictions to validate our adaptation model. We analyzed the effect on the adaptation dynamics when changing the adapted receptor activity, as well as introducing a non-regulatable CheB mutant to remove the negative feedback from phosphorylation of CheB by the kinase CheA. In both cases, our model is consistent with experimental measurements (Fig.~\ref{fig:dA_flowCheB_effects}), supporting the finding of multiple layers of feedback regulation in adaptation.

%other models
While the MWC model is relatively well established~\cite{SouBerg04,MelTu05,KeyEndSko06,EndWin06,EndSouWin2008}, we also considered alternative models for receptor signaling. These include a phase-separation model with mixed complexes separating into homogeneous complexes of Tar and Tsr at high ambient concentrations, as well as a lattice model with finite coupling between neighboring receptors (see Supplementary Text S1). Lattice models were previously suggested~\cite{DukeBray99, MelTu03b}, including a lattice formed by coupled CheA molecules~\cite{GoldLevBray09}, but were found to be inconsistent with FRET data~\cite{SkoEndWin06}.
We found that the dynamic MWC model describes dose-response curves far better than the alternative receptor signaling models investigated, particularly the reduced response amplitudes upon removal of attractant.
Furthermore, the data collapse we introduced in this paper enabled us to compare  different adaptation models proposed in the literature with FRET time-course data (Fig.~\ref{fig:compareadaptationmodels}). We found that while several models are consistent with the data, our model compared most favorably with the data.

% shortcomings
We chose a simple model for adaptation with very few fitting parameters to explain the observed asymmetry in adaptation time-courses, i.e. slow adaptation to addition and fast adaptation to removal of attractant.
Compared to the static MWC model, there are minor discrepancies between our model and experimental addition dose-response curves (Fig.~\ref{fig:DR_MWC_and_adaptation}). However, these can be rectified by refitting the dynamic MWC model by adjusting adaptation rates and receptor complex size simultaneously (see Supplementary Text S1), or by choosing an adaptation model with a more complex activity dependence. 
It should also be noted that adaptation rates needed to accurately describe dose-response curves are larger than those found when fitting the adaptation dynamics to the data collapse.
This discrepancy may in part be due to using only a single set of experimental data for the data collapse, while dose-response curves were averaged over at least three sets. In addition, more complex processes not taken into account in our simple adaptation model, e.g. limited supply of substrate (methionine) for methylation, or the binding and unbinding kinetics of ligand, may be important for describing the dynamics.

% potential mechanism for adaptation model
Although our adaptation model is independent of biochemical details, our predicted fast demethylation at high activities may be due to cooperativity of two CheB-P molecules.
According to {\it in vitro} experiments, CheB-P binding to the carboxyl-terminus of chemoreceptors induces an allosteric activation of the receptor, increasing the demethylation rate~\cite{BarBarHaz02}. However, in contrast to CheR, CheB-P binds only weakly to the tether~\cite{BarBarHaz02}.
Reconciling these two observations, it is conceivable that two CheB-P molecules are necessary for efficient demethylation at high activities: one CheB-P molecule may bind to a tether to allosterically activate a group of receptors (assistance neighborhood), while another CheB-P molecule demethylates the receptors in the neighborhood. As two CheB-P molecules are not required to bind to the same receptor, this mechanism is not contradicted by the small number of CheB molecules in a cell.
An alternative, simpler mechanism for cooperativity is dimerization of CheB-P molecules, which, however, has not been observed experimentally~\cite{AnaGouSto98,KenSou09}.

% remaining open questions
Our adaptation model likely also applies to attractants other than MeAsp, since the dynamics of adaptation only depend on the activity of receptor complexes, independent of the details of external ligand concentration. According to the MWC model, different attractants (or their mixture) are integrated at the level of the free-energy of a receptor complex, which determines its activity.
However, the imprecision of adaptation we found in FRET time courses at large MeAsp concentrations is in contrast to earlier experiments, which showed that the frequency of tumbling adapts precisely to aspartate, but not serine~\cite{BergBrown72,AlonSurLei99}. 
The imprecision in adaptation to serine is readily explained if the number of Tsr receptors is larger than the number of Tar receptors per complex, since the available receptor methylation sites in a complex are more
quickly used up in response to serine binding to Tsr receptors~\cite{EndWin06, HanEndWin08}. However, the ratio of Tar and Tsr per complex is strongly dependent on the growth conditions, making a definitive conclusion difficult~\cite{Kal2010}.
%{\color{blue} The imprecision in adaptation to serine is readily explained by the larger number of Tsr receptors per receptor complex, which, when binding serine, use up available receptor methylation sites more quickly~\cite{EndWin06, HanEndWin08}.}
%
Future experiments may show if the imprecision observed in adaptation to MeAsp in FRET is reflected also in the tumbling frequency, or if imprecise adaptation is compensated for in order to yield perfect adaptation at the behavioral level.

% You may title this section "Methods" or "Models". 
% "Models" is not a valid title for PLoS ONE authors. However, PLoS ONE
% authors may use "Analysis" 
\section*{Materials and Methods}

\paragraph{Strains}
Two different wild-type strains of {\it E. coli} were used. Wild-type strain 1 (WT1) is VS104 $\Delta$(cheY cheZ) that expresses the FRET pair consisting of CheY-YFP (YFP; yellow fluorescent protein) and its phosphatase CheZ-CFP (CFP; cyan fluorescent protein) from a pTrc-based plasmid pVS88, which carries pBR replication origin and ampicillin resistance and is inducible by isopropyl $\beta$-D-thiogalactoside (IPTG)~\cite{SouBerg2002}. Wild-type strain 2 (WT2) is VS124 $\Delta$(cheB cheY cheZ) transformed with pVS88 and pVS91, which carries pACYC replication origin and chloramphenicol resistance and encodes wild-type CheB under control of pBAD promoter inducible by L-arabinose. The CheB-mutant strain is VS124 $\Delta$(cheB cheY cheZ) transformed with pVS88 and pVS97, which is identical to pVS91 except it encodes the non-regulatable CheB$^{\text{D56E}}$. The D56E mutation was introduced into CheB by PCR. It prevents CheB phosphorylation, but allows protein to retain basal level of activity. Cells were grown at 275~rpm in a rotary shaker to mid-exponential phase ($\text{OD}_{\text{600}}\!\approx\!0.48$) in tryptone broth~(TB) medium supplemented with $100 \, \mu\text{g/ml}$ ampicillin, $34 \, \mu\text{g/ml}$ chloramphenicol, and $50 \, \mu\text{M}$ IPTG. WT and CheB mutant strains were induced by 0 and 0.0003\% arabinose, respectively, to produce comparable kinase activity (as assessed by the change in the level of FRET upon saturating stimulation). The CheB protein level was estimated using Western blots with CheB antibodies, and was at approximately 0.5-fold (WT2) and approximately 5-fold (CheB$^{\text{D56E}}$ mutant) the native level of CheB.

\paragraph{FRET measurements}
The experimental procedures follow those detailed by Sourjik and Berg~\cite{SouBerg2002}. 
Cells were tethered to a cover slip, and placed in a flow chamber. 
Cells were subject to a constant fluid flow of buffer or MeAsp at indicated concentration (flow speeds $1000\,\mu\text{l/min}$ for WT1, and $500\,\mu\text{l/min}$ for WT2 and CheB mutant, respectively).
Concentration step changes were achieved by abruptly switching between buffer and MeAsp, or different MeAsp concentrations. 
Fluorescence resonance energy transfer (FRET) between excited donor, CheZ-CFP, and acceptor, phosphorylated CheY-YFP, in a population of 300-500 cells was monitored using an epifluorescence microscopy setup. Emission light from CFP and YFP was collected and their intensity ratio $R$ was used to calculate the time-dependent number of interacting FRET pairs of CheZ-CFP and phosphorylated CheY-YFP in the cell population, which reflects the intracellular kinase activity~\cite{SouBerg2002}.
The number of FRET pairs normalized by its adapted pre-stimulus value (after adaptation to the ambient concentration, but before concentration step changes) was calculated from the ratio $R$ according to $(R - R_0)/[(\Delta Y/\Delta C) - R]/\{(R_{\text{pre}} - R_0)/[(\Delta Y/\Delta C) - R_{\text{pre}}]\}$~\cite{SouBerg2002}. The parameters $R_0$ and $R_{\text{pre}}$ are the ratio for a saturating dose of attractant and the pre-stimulus value, respectively, both of which are measured in each experiment. The fluorescence efficiency ratio $\Delta Y/\Delta C$ is determined by the experimental setup~\cite{SouVakBerg}, and was 0.43 ($\Delta Y/\Delta C\!\!=\!\!2.3$) for WT1 (WT2 and CheB mutant) experiments. 
FRET measurements were taken with a time resolution of 0.2~s (1~s) for WT1 (WT2 and CheB mutant).

\paragraph{Static MWC model} This model describes the response of adapted mixed receptor complexes to instantaneous MeAsp concentration step changes~\cite{SouBerg04,MelTu05,KeyEndSko06}.
According to this model, the activity of a mixed receptor complex is given by~$A=[1+\exp(F)]^{-1}$, where
\begin{equation}
 F = N \left[ \epsilon(m) + \nu_a \ln\left( \frac{1+c/K^{\text{off}}_a}{1+c/K^{\text{on}}_a}\right) + \nu_s \ln\left( \frac{1+c/K^{\text{off}}_s}{1+c/K^{\text{on}}_s}\right) \right]\label{eq:DF}
\end{equation}
is the free-energy difference between the {\it on} and {\it off} states of the complex.
The indexes $a$ and $s$ denote Tar and Tsr receptor, respectively. We assumed fractions of Tar and Tsr in a complex according to their wild-type ratio, $\nu_a\!\!:\!\!\nu_s\!\!=\!\!1\!\!:\!\!1.4$. The ligand dissociation constants for MeAsp are $K^{\text{on}}_a\!\!=\!\!0.5\,\text{mM}$, $K^{\text{off}}_a\!\!=\!\!0.02\,\text{mM}$, $K^{\text{on}}_s\!\!=\!\!10^6\,\text{mM}$, and $K^{\text{off}}_s\!\!=\!\!100\,\text{mM}$ \cite{KeyEndSko06}. The free-energy contribution $\epsilon(m)$ is attributed to methylation, and was recently extracted from dose-response curves for mutants resembling fixed methylation states \cite{EndSouWin2008}. We used the interpolating function $\epsilon(m)\!\!=\!\!1\!\!-\!\!0.5 m$ (for data and fit see {\it Inset} of Fig.~\ref{fig:methods}A). All energies are measured in units of $k_B T$ ($k_B$ being the Boltzmann constant and $T$ the absolute temperature).
Exponential rate constants for the ligand flow were obtained from fits to ligand flow profiles (cf.~Fig.~\ref{fig:methods}B), with $\lambda_{\text{add}}\!\!=\!\!0.6\,\text{s}^{-1}$ and $\lambda_{\text{rem}}\!\!=\!\!0.5\,\text{s}^{-1}$ for flow speed $1000 \, \mu\text{l/min}$, and $\lambda_{\text{add}}\!\!=\!\!0.27\,\text{s}^{-1}$ and $\lambda_{\text{rem}}\!\!=\!\!0.28\,\text{s}^{-1}$ for flow speed $500\, \mu\text{l/min}$.
The receptor complex size $N$ was estimated from least-squares fits to individual addition dose-response curves corresponding to specific ambient concentrations (and therefore adapted methylation levels). Note that complex size for removal may be different for each data point as cells are adapted to the increased concentration after each step change. The complex size grows with ambient concentration \cite{MelTu07,EndSouWin2008} in a roughly linear fashion, $N(c_0)\!\!=\!\!a_0+a_1 c_0$ with $a_0\!\!=\!\!17.5$ and $a_1\!\!=\!\!3.35 \, \text{mM}^{-1}$. Both, individually fitted $N$ values, as well as the fitting function $N(c_0)$, are shown in Fig.~\ref{fig:methods}A.
We assumed an adapted receptor complex activity $A^*\!\!=\!\!1/2.9\!\!\approx\!\!0.34$ for WT1 as assessed from the maximum and minimum values of the experimental dose-response data in Fig.~\ref{fig:DR_MWC_and_adaptation}. Steady-state activities for WT2 and CheB mutant were estimated to be $A^*\!\!\approx\!\!1/2$. For comparison of model and data, we normalized the receptor-complex activity for WT1, WT2 and CheB mutant by their respective activities when adapted to ambient concentration.

\paragraph{Precise adaptation}%
The dynamic MWC model combines the static MWC model with a dynamical model for adaptation.
In our model for precise adaptation, the rate of change of the average receptor methylation level $m$ is given by (Eq.~\ref{eq:preciseadaptation})
\begin{equation}
 \frac{dm}{dt}=g_R (1-A) - g_B A^3.\nonumber
\end{equation}
The methylation and demethylation rates do not depend directly on the concentration of MeAsp or the methylation level, only on the receptor complex activity $A$. Such dynamics leads to precise adaptation of the receptor complex activity to adapted level $A^*$ for a constant MeAsp stimulus~\cite{BarLei97,EndWin06}.
This model takes into account the receptor-activity dependence of the methylation and demethylation rates, as well as additional layers of feedback regulation for demethylation by CheB, including the activation of demethylation enzyme CheB by phosphorylation and potential cooperativity between phosphorylated CheB molecules.
For Fig.~\ref{fig:DR_MWC_and_adaptation}-\ref{fig:dAdm}, we determined the demethylation rate $g_B\!\!=\!\!0.11\,\text{s}^{-1}$ from a least-squares fit to addition and removal dose-response curves (WT1) using the receptor complex size $N(c_0)$ from the static MWC model. The methylation rate $g_R\!\!=\!\!0.0069\,\text{s}^{-1}$ is given by the condition that at steady state ($dm/dt\!\!=\!\!0$) the activity equals $A^*$.
The fit to the data collapse in Fig.~\ref{fig:compareadaptationmodels} resulted in $g_R\!\!=\!\!0.0019\,\text{s}^{-1}$ (and $g_B\!\!=\!\!0.030\,\text{s}^{-1}$), used in Fig.~\ref{fig:compareadaptationmodels} and \ref{fig:dA_flowCheB_effects} for WT1.
For WT2 in Fig.~\ref{fig:dA_flowCheB_effects}A, we used the same methylation rate constant as for WT1, but adjusted the demethylation rate constant to account for the increased adapted activity $A^*$.
%
% CheB mutant adaptation model
For the CheB mutant in Fig.~\ref{fig:dA_flowCheB_effects}C, the rate of change of the average receptor methylation level~$m$ is predicted to be
\begin{equation}
 \frac{dm}{dt}=g_R (1-A) - \tilde{g}_B A\label{eq:CheBadaptation},
\end{equation}
where we assume that the methylation rate is the same as for wild-type cells.
The demethylation rate constant $\tilde{g}_B\!\!=\!\!g_B {A^*}^2\!\!=\!\!g_B/4$ includes the basal activity of non-phos\-pho\-rylat\-able CheB. Hence, the only dependence of the demethylation rate on receptor complex activity is due to binding of CheB to active receptors.

\paragraph{Imprecise adaptation}%
We incorporate the effect of imprecise adaptation, as suggested by time courses (cf.~Fig.~\ref{fig:methods}C), by making methylation and demethylation rates for wild-type cells (WT1) depend on the methylation level~\cite{HanEndWin08}
\begin{equation}
 \frac{dm}{dt}=g_R \frac{m_{\text{max}}-m}{m_\text{max}-m+K}(1-A) - g_B \frac{m}{m+K}A^3\label{eq:impreciseadaptation}.
\end{equation}
The parameter $m_{\text{max}}$ is the maximum number of methylation sites per receptor, $K$ is the lower bound for the number of sites, which need to be available for efficient methylation or demethylation. We use $m_{\text{max}}\!\!=\!\!4.1$ to only allow Tar (not Tsr) receptors to become methylated (the total number of methylation sites of a receptor homodimer being 8). Further, we use $K\!\!=\!\!0.5$ to implement reduced efficiency of methylation or demethylation at a low number of available sites.
Figure~\ref{fig:methods}C shows time courses for adaptation to two concentration step changes using the precise and imprecise adaptation model ($g_R$ and $g_B$ are the same in both models). The imprecise adaptation model fits the time courses shown far better. However, there is a large variability of imprecision seen in different data sets and more experiments are needed to produce a general model of imprecise adaptation.

\paragraph{Rate of activity change}
To calculate the rate of activity change, the time courses for adaptation to step concentration changes were smoothed by averaging every 20 subsequent data points starting approximately 10 s after the step onset.
The derivative $dA/dt$ was approximated by the difference quotient.

% Do NOT remove this, even if you are not including acknowledgments
\section*{Acknowledgments}

We thank William Ryu, Thomas Shimizu, Yigal Meir and Ned Wingreen for helpful discussions. We also thank Sonja Schulmeister for help with CheB quantification. % D.C. and R.G.E. were supported by Biotechnological and Biological Sciences Research Council grant BB/G000131/1 and the Centre for Integrated Systems Biology at Imperial College. V.S. was supported by Deutsche Forschungsgesellschaft grant SO 421/3-3.

% \section*{References}
% \bibliographystyle{plos2009}
% \bibliography{maintext}

\begin{thebibliography}{}

\bibitem{JaasJackKea07} Jaasma MJ, Jackson WM, Tang RY, Keaveny TM (2007) Adaptation of cellular mechanical behavior to mechanical loading for osteoblastic cells. J Biomech 40: 1938-1945.

\bibitem{HillApiSchaf05} Hilliard MA, Apicella AJ, Kerr R, Suzuki H, Bazzicalupo P, et al. (2005) {\it In vivo} imaging of {\it C.~elegans} ASH neurons: cellular response and adaptation to chemical repellents. EMBO J 24: 63-72.

\bibitem{MarBibOes95} Marwan W, Bibikov SI, Montrone M, Oesterhelt D (1995) Mechanism of photosensory adaptation in {\it Halobacterium salinarium}. J Mol Biol 246: 493-499.

\bibitem{MuzzGomOuden09} Muzzey D, G\'{o}mez-Uribe CA, Mettetal JT, {van Oudenaarden} A (2009) A systems-level analysis of perfect adaptation in yeast osmoregulation. Cell 138: 160-171.

\bibitem{ShiZus94} Shi W, Zusman DR (1994) Sensory adaptation during negative chemotaxis in {\it Myxococcus xanthus}. J Bacteriol 176: 1517-1520.

\bibitem{SpehrHagZuf09} Spehr J, Hagendorf S, Weiss J, Spehr M, Leinders-Zufall T, et al. (2009) Ca$^{2+}$-calmodulin feedback mediates sensory adaptation and inhibits pheromone-sensitive ion channels in the vomeronasal organ. J Neurosci 29: 2125-2135.

\bibitem{ZigSull97} Zigmond SH, Sullivan SJ (1979) Sensory adaptation of leukocytes to chemotactic peptides. J Cell Biol 82: 517-527.

\bibitem{PughNikLam99} Pugh EN Jr, Nikonov S, Lamb TD (1999) Molecular mechanisms of vertebrate photoreceptor light adaptation. Curr Opin Neurobiol 9: 410-418.

\bibitem{BakWolStock06} Baker MD, Wolanin PM, Stock JB (2006) Systems biology of bacterial chemotaxis. Curr Opin Microbiol 9: 187-192.

\bibitem{Berg00} Berg HC (2000) Motile behavior of bacteria. Phys Today 53: 24-29.

\bibitem{FalHaz01} Falke JJ, Hazelbauer GL (2001) Transmembrane signaling in bacterial chemoreceptors. Trends Biochem Sci 26: 257-265.

\bibitem{Sou04} Sourjik V (2004) Receptor clustering and signal processing in {\it E. coli} chemotaxis. Trends Microbiol 12: 569-576.

\bibitem{WadArm04} Wadhams GH, Armitage JP (2004) Making sense of it all: bacterial chemotaxis. Nat Rev Mol Cell Biol 5: 1024-1037.

\bibitem{ClarkGrant05} Clark DA, Grant LC (2005) The bacterial chemotactic response reflects a compromise between transient and steady-state behavior. Proc Natl Acad Sci U S A 102: 9150-9155.

\bibitem{VlaLovSou08} Vladimirov N, L\o{}vdok L, Lebiedz D, Sourjik V (2008) Dependence of bacterial chemotaxis on gradient shape and adaptation rate. PLoS Comput Biol 4: e1000242.

\bibitem{VladSour09} Vladimirov N, Sourjik V (2009) Chemotaxis: how bacteria use memory. Biol Chem 390: 1097-1104.

\bibitem{EmoClu08} Emonet T, Cluzel P (2008) Relationship between cellular response and behavioral variability in bacterial chemotaxis. Proc Natl Acad Sci U S A 105: 3304-3309.

\bibitem{ZonBray09} Zonia L, Bray D (2009) Swimming patterns and dynamics of simulated {\it Escherichia coli} bacteria. J R Soc Interface 6: 1035-1046.

\bibitem{AndYiIgl06} Andrews BW, Yi TM, Iglesias PA (2006) Optimal noise filtering in the chemotactic response of Escherichia coli. PLoS Comput Biol 2: e154.

\bibitem{TuShimBerg08} Tu Y, Shimizu TS, Berg HC (2008) Modeling the chemotactic response of {\it Escherichia coli} to time-varying stimuli. Proc Natl Acad Sci U S A 105: 14855-14860.

\bibitem{BrayLevMorFir98} Bray D, Levin MD, Morton-Firth CJ (1998) Receptor clustering as a cellular mechanism to control sensitivity. Nature 393: 85-88.

\bibitem{KenSou09} Kentner D, Sourjik V (2009) Dynamic map of protein interactions in the {\it Escherichia coli} chemotaxis pathway. Mol Syst Biol 5: 238.

\bibitem{SouBerg2002} Sourjik V, Berg HC (2002) Receptor sensitivity in bacterial chemotaxis. Proc Natl Acad Sci U S A 99: 123-127.

\bibitem{SouBerg04} Sourjik V, Berg HC (2004) Functional interactions between receptors in bacterial chemotaxis. Nature 428: 437-441.

\bibitem{ChalWeis05} Chalah A, Weis RM (2005) Site-specific and synergistic stimulation of methylation on the bacterial chemotaxis receptor Tsr by serine and CheW. BMC Microbiol 5: 12.

\bibitem{Kosh81} Koshland DE (1981) Biochemistry of sensing and adaptation in a simple bacterial system. Ann Rev Biochem 50: 765-782.

\bibitem{LiStock09} Li Z, Stock JB  (2009) Protein carboxyl methylation and the biochemistry of memory. Biol Chem 390: 1087-1096.

\bibitem{AlonSurLei99} Alon U, Surette MG, Barkai N, Leibler S (1999) Robustness in bacterial chemotaxis. Nature 397: 168-171.

\bibitem{BergBrown72} Berg HC, Brown DA (1972) Chemotaxis in {\it Escherichia coli} analysed by three-dimensional tracking. Nature 239: 500-504.

\bibitem{LaiBarBar06} Lai WC, Barnakova LA, Barnakov AN, Hazelbauer GL (2006) Similarities and differences in interactions of the activity-enhancing chemoreceptor pentapeptide with the two enzymes of adaptational modification. J~Bacteriol 188: 5646-5649.

\bibitem{LiHaz06} Li M, Hazelbauer GL (2006) The carboxyl-terminal linker is important for chemoreceptor function. Mol~Microbiol 60: 469-479.

\bibitem{ToeGoyStock79} Toews ML, Goy MF, Springer MS, Adler J (1979)  Attractants and repellents control demethylation of methylated chemotaxis proteins in {\it Escherichia coli}. Proc Natl Acad Sci U S A 76: 5544-5548.

\bibitem{LiHaz05} Li M, Hazelbauer GL (2005) Adaptational assistance in clusters of bacterial chemoreceptors. Mol Microbiol 56: 1617-1626.

\bibitem{LiHaz04} Li M, Hazelbauer GL (2004) Cellular stoichiometry of the components of the chemotaxis signaling complex. J Bacteriol 186: 3687-3694.

\bibitem{SimStockStock87} Simms SA, Stock AM, Stock JB (1987) Purification and characterization of the S-adenosylmethionine: glutamyl methyltransferase that modifies membrane chemoreceptor proteins in bacteria. J Biol Chem 262: 8537-8543.

\bibitem{LaiHaz05} Lai WC, Hazelbauer GL (2005) Carboxyl-terminal extensions beyond the conserved pentapeptide reduce rates of chemoreceptor adaptational modification. J Bacteriol 187: 5115-5121.

\bibitem{KortGoyAdl75} Kort EN, Goy MF, Larsen SH, Adler J (1975) Methylation of a membrane protein involved in bacterial chemotaxis. Proc Nat Acad Sci U S A 72: 3939-3943.

\bibitem{ChelGutKosh84} Chelsky D, Gutterson NI, Koshland DE (1984) A diffusion assay for detection and quantitation of methyl-esterified proteins on polyacrylamide gels. Anal Biochem 141: 143-148.

\bibitem{WuLiWeis96} Wu J, Li J, Li G, Long DG, Weis RM (1996) The receptor binding site for the methyltransferase of bacterial chemotaxis is distinct from the sites of methylation. Biochemistry 35: 4984-4993.

\bibitem{BarBarHaz02} Barnakov AN, Barnakova LA, Hazelbauer GL (2002) Allosteric enhancement of adaptational demethylation by a carboxyl-terminal sequence on chemoreceptors. J~Biol~Chem 277: 42151-42156.

\bibitem{EndSouWin2008} Endres RG, Oleksiuk O, Hansen CH, Meir Y, Sourjik V, et al. (2008) Variable sizes of {\it Escherichia coli} chemoreceptor signaling teams. Mol Syst Biol 4: 211.

\bibitem{EndWin06} Endres RG, Wingreen NS (2006) Precise adaptation in bacterial chemotaxis through ``assistance neighborhoods''. Proc Natl Acad Sci U S A 103: 13040-13044.

\bibitem{KeyEndSko06} Keymer JE, Endres RG, Skoge M, Meir Y, Wingreen NS (2006) Chemosensing in {\it Escherichia coli}: two regimes of two-state receptors. Proc Natl Acad Sci U S A 103: 1786-1791.

\bibitem{MelTu05} Mello BA, Tu Y (2005) An allosteric model for heterogeneous receptor complexes: understanding bacterial chemotaxis responses to multiple stimuli. Proc Natl Acad Sci U S A 102: 17354-17359.

\bibitem{BarLei97} Barkai N, Leibler S (1997) Robustness in simple biochemical networks. Nature 387: 913-917.

\bibitem{MelTu03} Mello BA, Tu Y (2003) Perfect and near-perfect adaptation in a model of bacterial chemotaxis. Biophys J 84: 2943-2956.

\bibitem{YiHuaDoy00} Yi T, Huang Y, Simon MI, Doyle J (2000) Robust perfect adaptation in bacterial chemotaxis through integral feedback control. Proc Natl Acad Sci U S A 97: 4649-4653.

\bibitem{KolLovSou05} Kollmann M, L\o{}vdok L, Bartholome K, Timmer J, Sourjik V (2005) Design principles of a bacterial signalling network. Nature 438: 504-507.

\bibitem{HanEndWin08} Hansen CH, Endres RG, Wingreen NS (2008) Chemotaxis in {\it Escherichia coli}: a molecular model for robust precise adaptation. PLoS Comput Biol 4: e1.

\bibitem{UrsHuaPhil07} Ursell T,  Huang KC,  Peterson E,  Phillips R (2007) Cooperative Gating and Spatial Organization of Membrane Proteins through Elastic Interactions. PLoS Comput Biol 3: e81.

\bibitem{GurKahEnd09} Gurry T,  Kahramano\u{g}ullar\i,  Endres RG (2009) Biophysical Mechanism for Ras-Nanocluster Formation and Signaling in Plasma Membrane. PLoS ONE 4: e6148.

\bibitem{MelTu07} Mello BA, Tu Y (2007) Effects of adaptation in maintaining high sensitivity over a wide range of backgrounds for {\it Escherichia coli} chemotaxis. Biophys J 92: 2329-2337.

\bibitem{KalTuWu09} Kalinin YV, Jiang L, Tu Y, Wu M (2009) Logarithmic sensing in {\it Escherichia coli} bacterial chemotaxis. Biophys J 96: 2439-2448.

\bibitem{DukeBray99} Duke TAJ, Bray D (1999) Heightened sensitivity of a lattice of membrane receptors. Proc Natl Acad Sci U S A 96: 10104-10108.

\bibitem{MelTu03b} Mello BA, Tu Y (2003) Quantitative modeling of sensitivity in bacterial chemotaxis: The role of coupling among different chemoreceptor species. Proc Natl Acad Sci U S A 100: 8223-8228.

\bibitem{GoldLevBray09} Goldman JP, Levin MD, Bray D (2009) Signal amplification in a lattice of coupled protein kinases. Mol BioSyst 5: 1853-1859.

\bibitem{SkoEndWin06} Skoge ML, Endres RG, Wingreen NS (2006) Receptor-receptor coupling in bacterial chemotaxis: evidence for strongly coupled clusters. Biophys J 90: 4317-4326.

\bibitem{AnaGouSto98} Anand GS, Goudreau PN, Stock AM (1998) Activation of methylesterase CheB: evidence of a dual role for the regulatory domain. Biochemistry 37: 14038-14047.

\bibitem{Kal2010} Kalinin~Y, Neumann~S, Sourjik~V, Wu~M (2010) Responses of {\it Escherichia coli} bacteria to two opposing chemoattractant gradients depend on the chemoreceptor ratio. J Bacteriol 192: 1796-1800.

\bibitem{SouVakBerg} Sourjik V, Vaknin A, Shimizu TS, Berg HC (2007) {\it In vivo} measurement by FRET of pathway activity in bacterial chemotaxis. Methods Enzymol 423: 363-391.



\end{thebibliography}

\section*{Figure Legends}

% \section*{Tables}
%\begin{table}[!ht]
%\caption{
%\bf{Table title}}
%\begin{tabular}{|c|c|c|}
%table information
%\end{tabular}
%\begin{flushleft}Table caption
%\end{flushleft}
%\label{tab:label}
% \end{table}

\end{document}

\maketitle

\tableofcontents

\section{Review of chemotaxis signaling pathway}

The bacterium {\it Escherichia coli} chemotaxes by utilizing a biased random walk towards a nutrient source (or away from a toxin source)~\cite{Berg00,FalHaz01,Sou04,WadArm04}. The swimming path consists of runs, i.e. straight swimming driven by coherent motion of flagella, and tumbles characterized by lack of net movement and random reorientation of the cell.

\begin{figure}[t]
  \centering
    \includegraphics[]{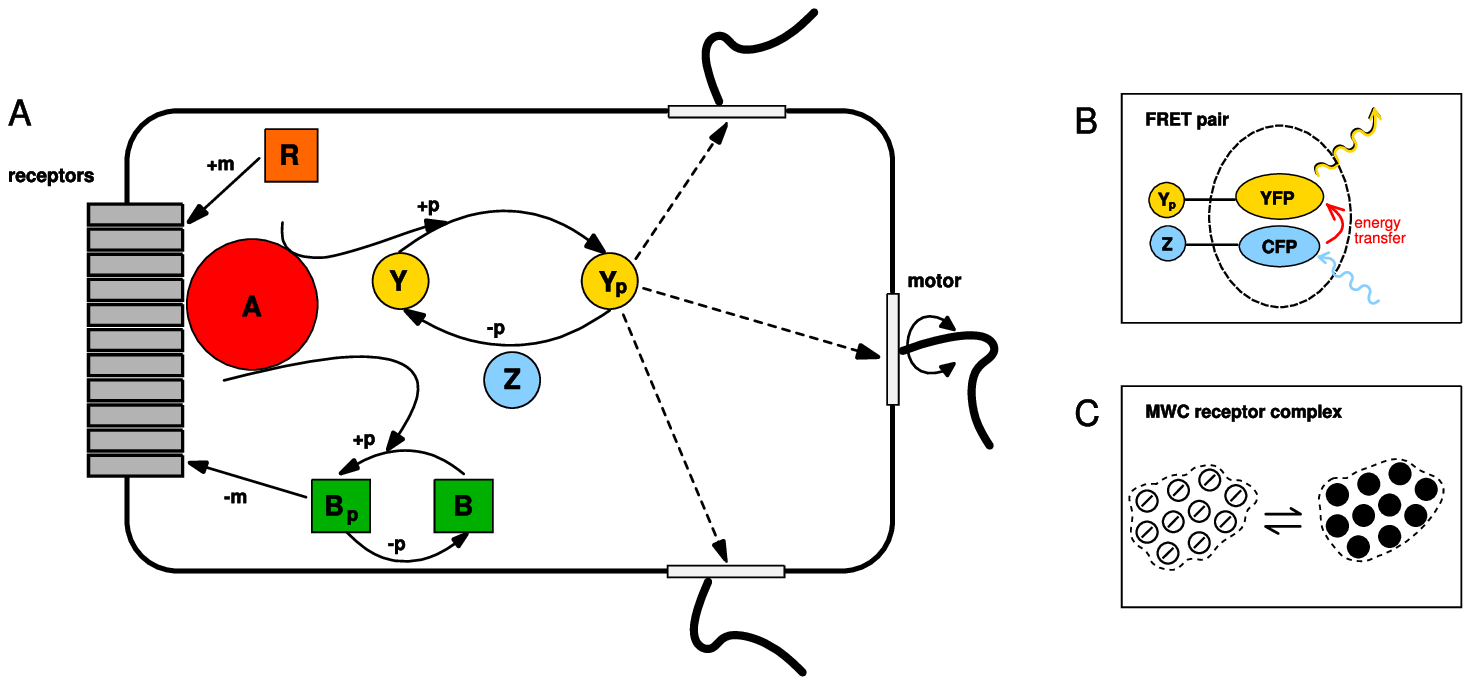}
\caption{\label{fig:pathway}Schematics of chemotaxis signaling and its measurement by FRET. (A) Chemotaxis signaling pathway in {\it E. coli} from receptors to rotary motors and flagella, including phosphotransfer from CheA to CheY and CheB, CheY-P diffusion to rotary motors, and dephosphorylation of CheY-P by phosphatase CheZ. Adaptation involves receptor methylation by CheR and demethylation by CheB-P. (B) FRET pair CFP/YFP used in experiments by Sourjik and Berg~\cite{SouBerg02a}; YFP tags CheY and CFP tags CheZ. CFP is excited by laser light and transfers its energy to YFP during FRET. YFP de-excites by fluorescence. (C) Fast switching between {\it on} (white discs) and {\it off} state (black discs) of an MWC receptor complex.}
\end{figure}

The molecular components of the chemotaxis signaling pathway and relationships between them are well-characterized~\cite{KenSou09}, and are shown schematically in Fig.~\ref{fig:pathway}. Transmembrane chemoreceptors localize predominantly at cell poles, where they form large clusters. There are five different types of chemoreceptors, each with specific sensing capabilities. The two most abundant receptor types, Tar and Tsr, bind respectively the amino acids aspartate (and its non-metabolizable analogue MeAsp) and serine. Tsr also binds aspartate and MeAsp with much lower affinity.
Binding of ligand induces signaling by the receptor across the membrane to the kinase CheA. CheA as well as the protein CheW (not shown in Fig.~\ref{fig:pathway}) have been suggested to be involved in receptor-receptor coupling and signal integration. When active, CheA autophosphorylates and rapidly passes on a phosphoryl group to its response regulators CheY and CheB. Phosphorylated CheY (CheY-P) diffuses to the rotary motors which drive the cell's flagella. Upon binding to the motors, CheY-P induces a switch in rotary direction resulting in tumbling.
CheZ is a phosphatase of CheY-P.
Attractant binding reduces the activity of CheA, lowering the concentration of CheY-P in the cell, and therefore suppressing tumbling. In contrast, repellents cause an increase of activity, enhancing tumbling.

Adaptation is mediated by the proteins CheR and CheB. CheR methylates receptors to enhance their signaling activity. Phosphorylated CheB (CheB-P) demethylates receptors, which reduces their activity. During persistent stimulation by a chemical, the combined effect of receptor methylation by CheR and demethylation by CheB-P leads to adaptation of the kinase activity to a steady-state, allowing the sensing of new changes in attractant or repellent concentrations.

\section{Whole-pathway model}

We consider the following reactions shown in Fig.~\ref{fig:pathway}: (1) auto-phosphorylation of CheA and formation of CheA-P (concentrations $[A_p]$) when receptors are active, (2) phosphorylation of CheY and formation of CheY-P ($[Y_p]$), (3) association of CheY-P and CheZ ($[Y_pZ]$), leading to the dephosphorylation of CheY-P and dissociation into CheY and CheZ, and (4) phosphorylation of CheB and formation of CheB-P ($[B_p]$).

\begin{figure}[t]
  \centering
    \includegraphics[]{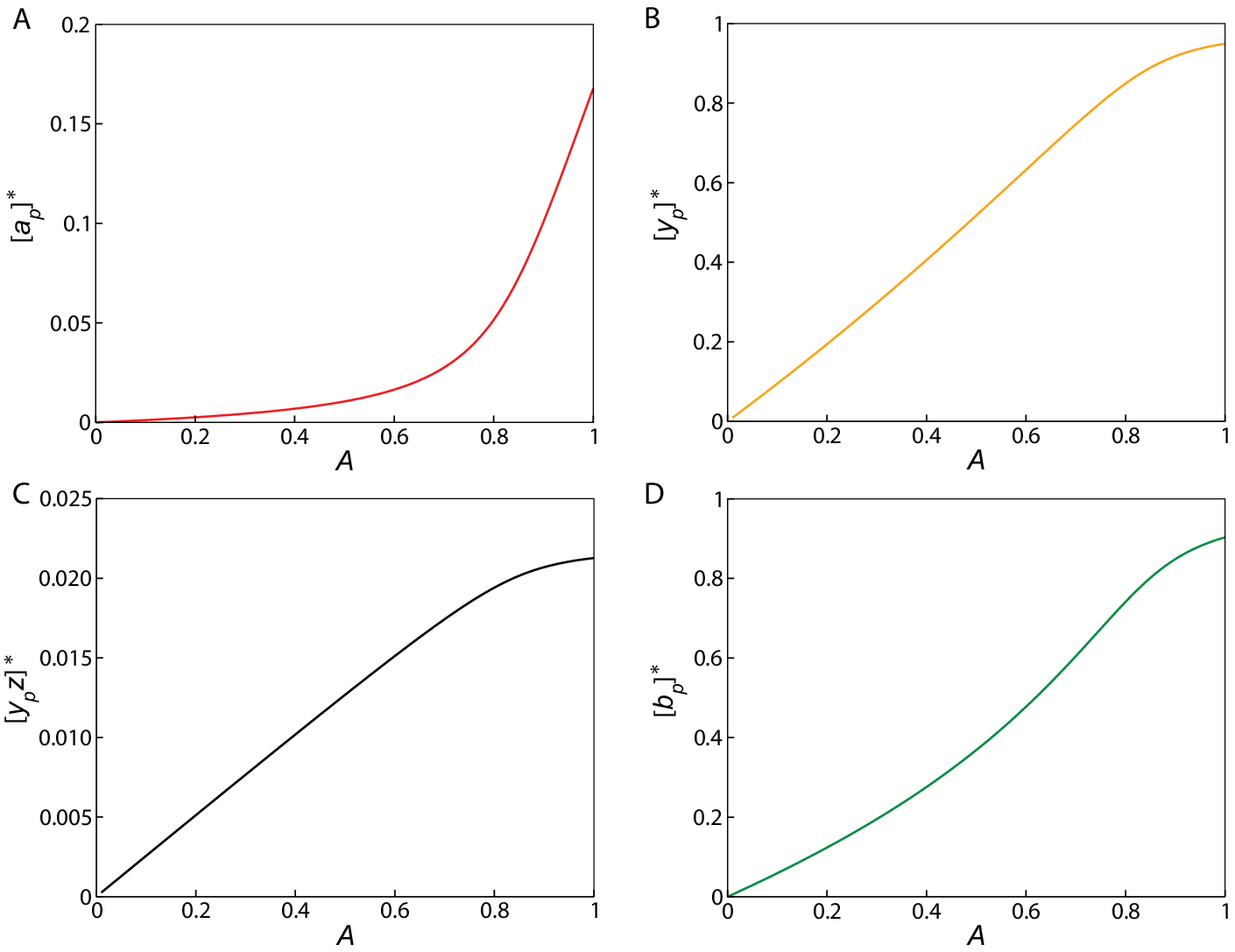}
\caption{\label{fig:steadystate}Steady-state concentrations of individual proteins and CheY-P/CheZ pairs for the whole-pathway model for Eq.~\ref{eq:wholepathway_Aresc}-\ref{eq:wholepathway_Bresc}, as a function of the receptor complex activity $A$. Note the different scales of the vertical axes.
The adapted activity is $A^*\approx$1/3.
}
\end{figure}

Assuming the law of mass-action, our model comprises the following set of ordinary differential equations:
\begin{eqnarray}
 \frac{d{[A_p]}}{dt}	& = & \underbrace{A \cdot k_A \left( [A]_{\text{tot}}-[A_p] \right)}_{\text{CheA autophosphorylation}} - \underbrace{k_{Y} \left( [Y]_{\text{tot}}-[Y_p] \right) [A_p]}_{\text{CheY phosphorylation}} - \underbrace{k_{B} \left( [B]_{\text{tot}}-[B_p] \right) [A_p]}_{\text{CheB phosphorylation}} \label{eq:wholepathway_A}
\end{eqnarray}
\begin{eqnarray}
 \frac{d{[Y_p]}}{dt} 	& = & \underbrace{k_{Y} \left( [Y]_{\text{tot}}-[Y_p] \right) [A_p]}_{\text{CheY phosphorylation}} - \underbrace{k_{1} \left([Z]_{\text{tot}}-[Y_pZ]\right) [Y_p]}_{\substack{\text{CheY-P/CheZ}\\\text{association}}} + \underbrace{k_2 [Y_pZ]}_{\substack{\text{CheY-P/CheZ}\\\text{dissociation}}} \label{eq:wholepathway_Y}\\
 \frac{d{[Y_pZ]}}{dt}	& = & \underbrace{k_1 ([Z]_{\text{tot}}-[Y_pZ]) [Y_p]}_{\text{CheY-P/CheZ association}} - \underbrace{\left(k_2+k_3\right) [Y_pZ]}_{\substack{\text{CheY-P/CheZ dissociation}\\\text{and CheY-P dephosphorylation}}} \label{eq:wholepathway_YZ}\\
 \frac{d{[B_p]}}{dt}	& = & \underbrace{k_B \left( [B]_{\text{tot}}-[B_p] \right)[A_p]}_{\text{CheB phosphorylation}} - \underbrace{k_{-B} [B_p],}_{\substack{\text{spontaneous}\\\text{dephosphorylation of CheB-P}}} \label{eq:wholepathway_B}
\end{eqnarray}
where the $k_i$ (with $i= 1,2,3,A,B,-B$ and $Y$) are kinetic rate constants for the individual reactions.
The activity $A$ of a receptor complex in Eq.~\ref{eq:wholepathway_A} is determined by the MWC model, given by
\begin{eqnarray}
A & = & \frac{1}{1+e^{F}},\,\,\text{with}\\
F & = & N \left[ \epsilon(m) + \nu_a \ln\left( \frac{1+c/K^{\text{off}}_a}{1+c/K^{\text{on}}_a}\right) + \nu_s \ln\left( \frac{1+c/K^{\text{off}}_s}{1+c/K^{\text{on}}_s}\right) \right],
\end{eqnarray}
as described in the main text.
In addition, we include the adaptation dynamics by methylation and demethylation of receptors, catalyzed by CheR and CheB-P, respectively (cf. Eq.~2 in the main text),%Eqref
\begin{eqnarray}
 \frac{d{m}}{dt} & = & \underbrace{g_R \left(1-A\right)}_{\text{methylation by Che-R}} - \underbrace{\hat{g}_B [B_p]^2 A,}_{\text{demethylation by CheB-P}}\label{eq:meth}\\
		& = & g_R (1-A) - g_B A^3\label{eq:meth_main}
\end{eqnarray}
where $g_R$ and $\hat{g}_B$ are effective rate constants, and $\hat{g}_B [B_p]^2\approx g_B A^2$ as $[B_p]$ is approximately proportional to the receptor complex activity (Fig.~\ref{fig:steadystate}D). This adaptation model is further explained in the next section.
The parameter values we used for the whole-pathway model are listed in Table~\ref{tab:wholepathway_parameters}.

\subsection{Rescaling of parameters}%

In order to reduce the number of parameters, we normalize the protein concentrations by their respective total concentrations in the cell, {${[A_p]} \rightarrow [a_p]=[A_p]/[A]_{\text{tot}}$}, {${[Y_p]} \rightarrow [y_p]=[Y_p]/[Y]_{\text{tot}}$}, {${[Y_pZ]} \rightarrow [y_pz]=[Y_pZ]/[Y]_{\text{tot}}$} and {${[B_p]} \rightarrow [b_p]=[B_p]/[B]_{\text{tot}}$}.
Furthermore, we rescale the time by the autophosphorylation rate of CheA, $k_A$, $t\rightarrow \tau=k_A \cdot t$, and introduce rescaled rate constants according to {$k_1 \rightarrow \kappa_1=k_1 [Y]_{\text{tot}}/k_A$}, {$k_2\rightarrow \kappa_2=k_2/k_A$}, {$k_3\rightarrow \kappa_3=k_3/k_A$}, {$k_Y \rightarrow \kappa_Y=k_Y [A]_{\text{tot}}/k_A$}, {$k_B \rightarrow \kappa_B=k_B [A]_{\text{tot}}/k_A$} and {$k_{-B} \rightarrow \kappa_{-B}=k_{-B}/k_A$}. 
Overall, this transformation yields dimensionless kinetic variables and parameters by measuring phosphorylated protein fractions in units of total protein concentrations and rate constants relative to the autophosphorylation rate constant of CheA. 
Using the ratios of total protein concentrations, {$\alpha_1 = [Y]_{\text{tot}}/[A]_{\text{tot}}$}, {$\alpha_2 = [B]_{\text{tot}}/[A]_{\text{tot}}$}, and {$\alpha_3 = [Z]_{\text{tot}}/[Y]_{\text{tot}}$}, we obtain the transformed set of equations
\begin{eqnarray}
 \frac{d}{d \tau}{[a_p]}	& = & A \cdot \left( 1-[a_p] \right)  - \alpha_1 \kappa_{Y} \left( 1-[y_p] \right) [a_p] - \alpha_2 \kappa_{B} \left( 1-[b_p] \right) [a_p] \label{eq:wholepathway_Aresc}\\
 \frac{d}{d \tau}{[y_p]} 	& = & \kappa_{Y} \left( 1-[y_p] \right) [a_p] - \kappa_{1} \left(\alpha_3-[y_pz]\right) [y_p] + \kappa_2 [y_pz] \label{eq:wholepathway_Yresc}\\
 \frac{d}{d \tau}{[y_pz]}	& = & \kappa_1 (\alpha_3-[y_pz]) [y_p] - \left(\kappa_2+\kappa_3\right) [y_pz] \label{eq:wholepathway_YZresc}\\
 \frac{d}{d \tau}{[b_p]}	& = & \kappa_B \left( 1-[b_p] \right)[a_p] - \kappa_{-B} [b_p] \label{eq:wholepathway_Bresc}.
\end{eqnarray}

The transformed equation for the methylation level of receptors is obtained by replacing time $t\rightarrow \tau$, $g_R\rightarrow \gamma_R=g_R/k_A$ and $\hat{g}_B\rightarrow\gamma_B=\hat{g}_B B_{\text{tot}}^2/k_A$ in Eq.~\ref{eq:meth}, yielding
\begin{equation}
 \frac{d{m}}{d\tau} = \gamma_R \left(1-A\right) - \gamma_B [b_p]^2 A\label{eq:methresc}.
\end{equation}
The new parameter values of this transformed model are listed in Table~\ref{tab:wholepathway_parameters}.

\subsection{Steady-state concentrations}

We analyzed the steady-state concentrations of phosphorylated proteins and CheY-P/CheZ pairs. Setting the time-derivatives of Eq.~\ref{eq:wholepathway_Aresc}-\ref{eq:wholepathway_Bresc} to zero, we solved for the steady-state concentrations of CheA-P, CheY-P and CheB-P, as well as the concentration of CheY-P/CheZ pairs as a function of the receptor complex activity $A$. The results are shown in Fig.~\ref{fig:steadystate}. CheA-P shows a strong non-linear dependence on the activity $A$, i.e., it is strongly activated at high receptor complex activity. It is also notable that only a small fraction of the CheA concentration is phosphorylated at maximal receptor activity $A=1$, which nicely fits estimates from {\it in vitro} measurements~\cite{WolBakStock06}.
All other phosphorylated fractions of protein, as well as the concentration of CheY-P/CheZ pairs are approximately proportional to receptor complex activity $A$.
\begin{table}
\caption{\label{tab:wholepathway_parameters}
Parameters of the whole-pathway model for chemotaxis signaling for Eq.~\ref{eq:wholepathway_A}-\ref{eq:meth}, including references to literature values where possible, and rescaled parameters for Eq.~\ref{eq:wholepathway_Aresc}-\ref{eq:methresc}.
The literature values are given in parentheses where different from our parameter values. $k_{1}$ was determined by the condition that at steady-state with $A^*$=1/3, the concentration $[Y_p]^*=[Y]_{\text{tot}}/3$~\cite{SouBerg02b}. $g_R$ was determined by the steady-state activity $A^*$ and the value for $\hat{g}_B$.}
\begin{center}
\begin{tabular}{|p{1.8cm}|l|p{4.5cm}||p{1.8cm}|l|}
\hline
Parameter			& Value						& Reference			& Scaled \mbox{parameter}	& Value	\\
\hline
$[A]_{\text{tot}}$ 		& 5 $\mu$M					& \cite{SouBerg02b}		& $\alpha_1$	 		& 1.94	\\
$[B]_{\text{tot}}$		& 0.28 $\mu$M					& \cite{LiHaz04}		& $\alpha_2$			& 0.056	\\
$[Y]_{\text{tot}}$ 		& 9.7 $\mu$M					& \cite{LiHaz04}		& $\alpha_3$ 			& 0.113	\\
$[Z]_{\text{tot}}$	 	& 1.1 $\mu$M					& \cite{SouBerg02b}		& $\kappa_Y$ 			& 50 \\
$k_A$		 		& 10 s$^{-1}$					& \cite{WolBakStock06}		& $\kappa_B$ 			& 7.5 \\
$k_Y$ 				& 100 $\mu$M$^{-1}$ s$^{-1}$			& \cite{StewJahrPar00}		& $\kappa_1$	 		& 4.88	\\
$k_B$ 				& 15 $\mu$M$^{-1}$ s$^{-1}$ 			& \cite{StewJahrPar00}		& $\kappa_2$		 	& 0.05	\\
$k_1$ 				& 5.0 $\mu$M$^{-1}$ s$^{-1}$			& \cite{SouBerg02b}		& $\kappa_3$ 			& 20	\\
$k_2$ 				& 0.5 s$^{-1}$ 					& \cite{SouBerg02b}		& $\kappa_{-B}$ 		& 0.135	\\
$k_3$ 				& 200 s$^{-1}$					& (30 s$^{-1}$) \cite{SouBerg02b} & $\gamma_R$			& 0.0006	\\
$k_{-B}$			& 1.35 s$^{-1}$					& (0.35 s$^{-1}$) \cite{BrayBou95,Stew93} & $\gamma_B$		& 0.0246	\\
$g_R$				& 0.006	s$^{-1}$				& --- & & \\
$\hat{g}_B$			& 3.14 $\mu$M$^{-2}$ s$^{-1}$			& --- & &\\
\hline
\end{tabular}
\end{center}
\end{table}
\begin{figure}[htp]
  \centering
    \includegraphics[width=0.7\textwidth]{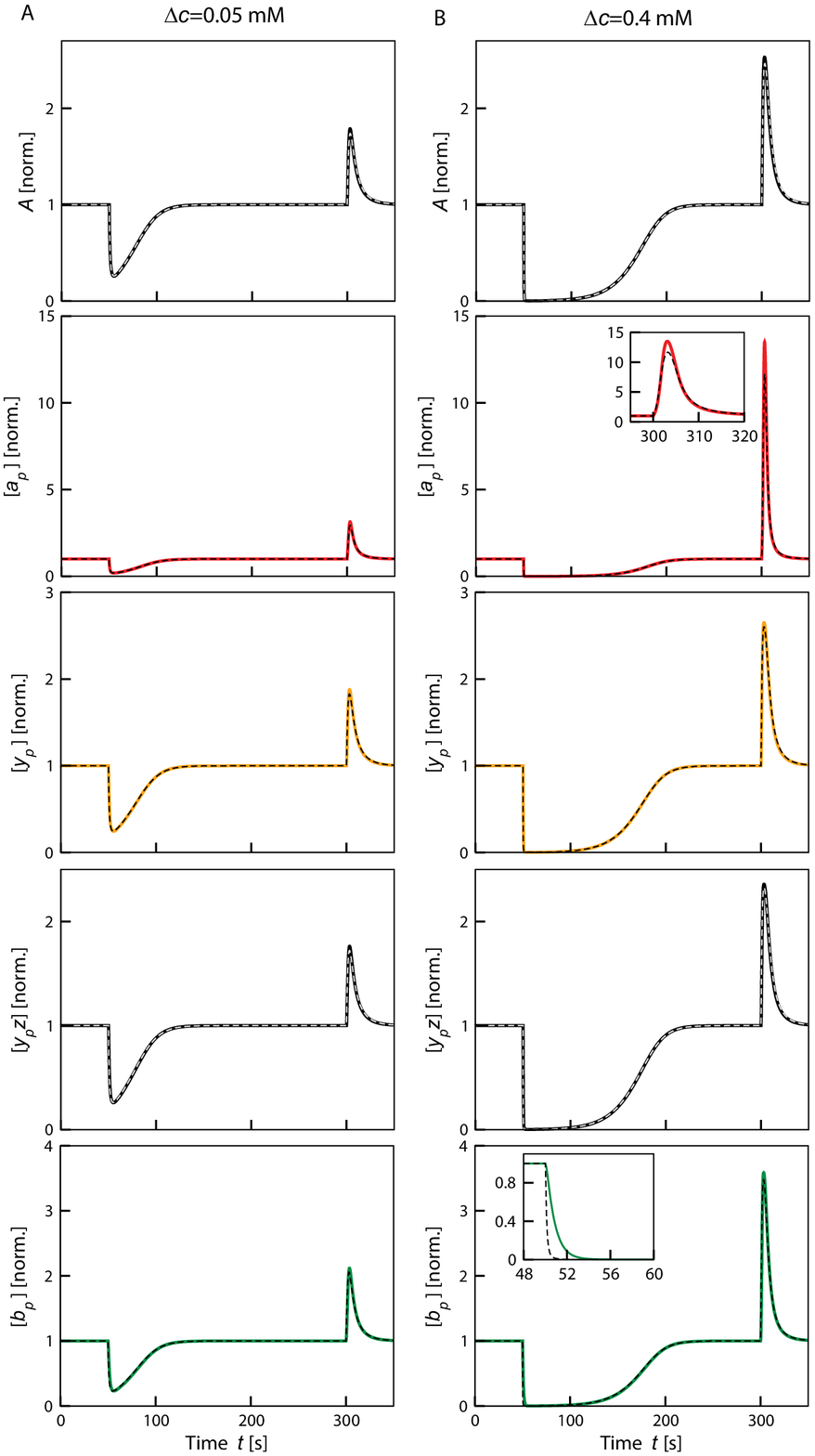}
\caption{\label{fig:wholepathway_step}Time courses for the concentrations of phosphorylated proteins and CheY-P/CheZ pairs according to the whole-pathway model Eq.~\ref{eq:wholepathway_Aresc}-\ref{eq:methresc}, using ambient MeAsp concentration $c_0$=0.1 mM and two different concentration step sizes, $\Delta c$=0.05 mM (column A) and $\Delta c$=0.4 mM (column B). Thick solid curves represent the model with parameter values as in table~\ref{tab:wholepathway_parameters}, thin dashed lines assume the quasi steady-state for all phosphorylation reactions (phosphorylation and dephosphorylation kinetic constants in the model increased by a factor 20). Insets zoom into the dynamics around the addition or removal time, respectively. Concentrations were normalized to their respective adapted values.
}
\end{figure}

\subsection{Time courses and steady-state assumption}

We tested if the phosphorylation and CheY-P/CheZ association reactions, Eq.~\ref{eq:wholepathway_Aresc}-\ref{eq:wholepathway_Bresc}, are in quasi-steady state compared to the slower methylation and demethylation reactions of receptors, Eq.~\ref{eq:methresc}. For this purpose, we increased all rate constants for phosphorylation, dephosphorylation, as well as CheY-P/CheZ association and dissociation by one order of magnitude, such that concentrations are forced to be in quasi steady-state at each time point.
Comparing the results to the time courses with the original parameter values shown in Fig.~\ref{fig:wholepathway_step}, we found only minor deviations (exemplified in insets).
Therefore, the above mentioned reactions are indeed in quasi-steady state to a good approximation.
This, together with the approximate linearity of the steady-state concentration of CheY-P/CheZ pairs as function of receptor complex activity $A$, permits us to replace the number of FRET (CheY-P/CheZ) pairs by the receptor complex activity (with appropriate proportionality factors) as assumed in the main text. Similarly, Eq.~2 %Eqref
in the main text arises by replacing CheB-P concentration in the demethylation rate in Eq.~\ref{eq:methresc} by the receptor complex activity $A$, where the methylation and demethylation rate constants are $g_R=\gamma_R k_A$ and $g_B=\gamma_B k_A \left([b_p]/A\right)^2\approx\gamma_B k_A$, respectively.

\section{Additional data and best fit of dynamic MWC model}

\begin{figure}[t]
  \centering
    \includegraphics[]{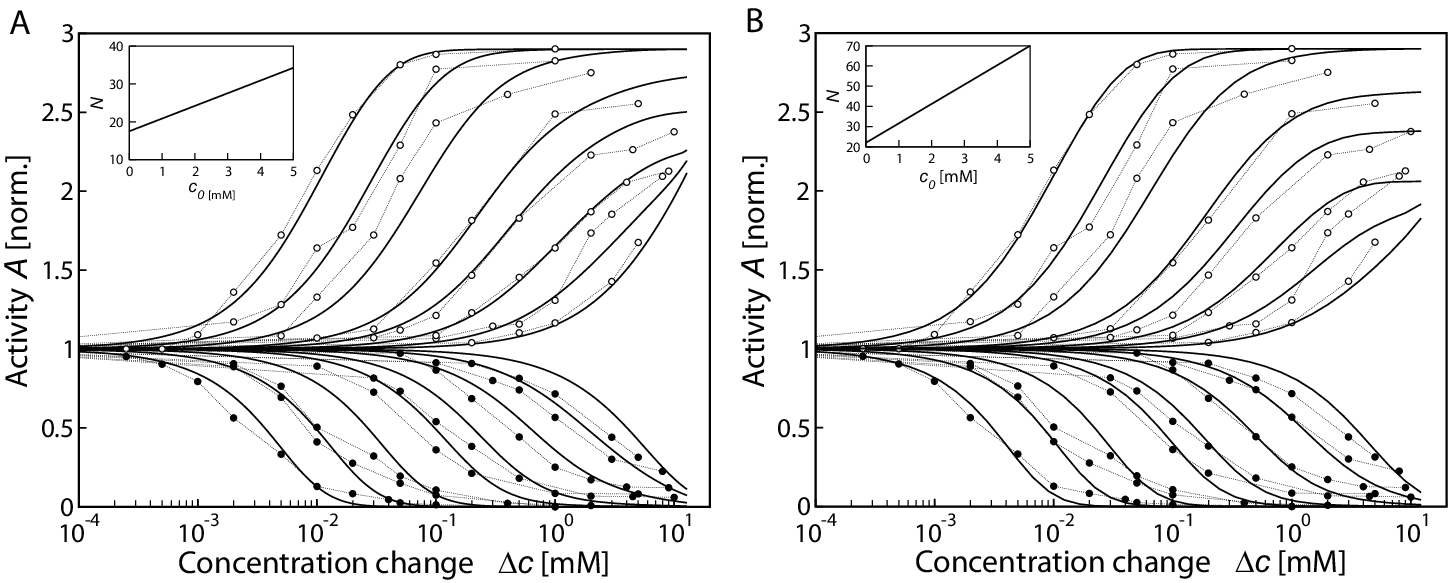}
\caption{\label{fig:DR_bestfit}Dynamic MWC model. (A) Same as Fig.~1 in the main text (main panel), however showing additional, previously unpublished data. Shown are dose-response curves for wild-type cells to step changes of MeAsp concentration (after adaptation to ambient concentrations 0, 0.03, 0.1, 0.3, 0.5, 1, 2 and 5 mM). Symbols represent averaged response from FRET data as measured by Sourjik and Berg~\cite{SouBerg02a}. Filled and open circles correspond to response to addition and removal of attractant, respectively. Solid lines represent the dynamic MWC model of mixed Tar/Tsr-receptor complexes, including ligand rise (addition) and fall (removal), as well as adaptation (receptor methylation) dynamics. (B) Best global fit of dynamic MWC model with fitting parameters $g_B$=0.127 s$^{-1}$, $K^{\text{off}}_a$=0.02 mM, $K^{\text{on}}_a$=0.50 mM,  $K^{\text{off}}_s$=216 mM, $K^{\text{on}}_s$=10$^6$ mM, as well as $a_0=22$ and $a_1$=9.6 mM$^{-1}$ for the total receptor complex size $N=a_0+a_1 c_0$.}
\end{figure}

In Fig.~\ref{fig:DR_bestfit} we show additional, previously unpublished dose-response data measured as described in the main text (cf. Fig.~1 in the main text). The model in panel~A is the dynamic MWC model from the main text.
Panel~B shows the best fit of the dynamic MWC model, where we used the demethylation rate constant $g_B$, the coefficients determining the receptor complex size as a linear function of ambient concentration, and the ligand dissociation constants of Tar and Tsr as fitting parameters. We found that parameters overall stay similar to the previously used parameters; in particular the ligand dissociation constants do not change significantly. The main difference is larger receptor complex sizes than determined by fitting the static MWC model to individual addition dose-response curves. To compensate for the larger complex sizes, the adaptation rates are also slightly increased, marking the trade-off between increased activity responses by larger complex sizes and reduced activity responses by faster adaptation (controlling for MeAsp concentration dynamics).
\begin{figure}[t]
  \centering
    \includegraphics[width=0.5\textwidth]{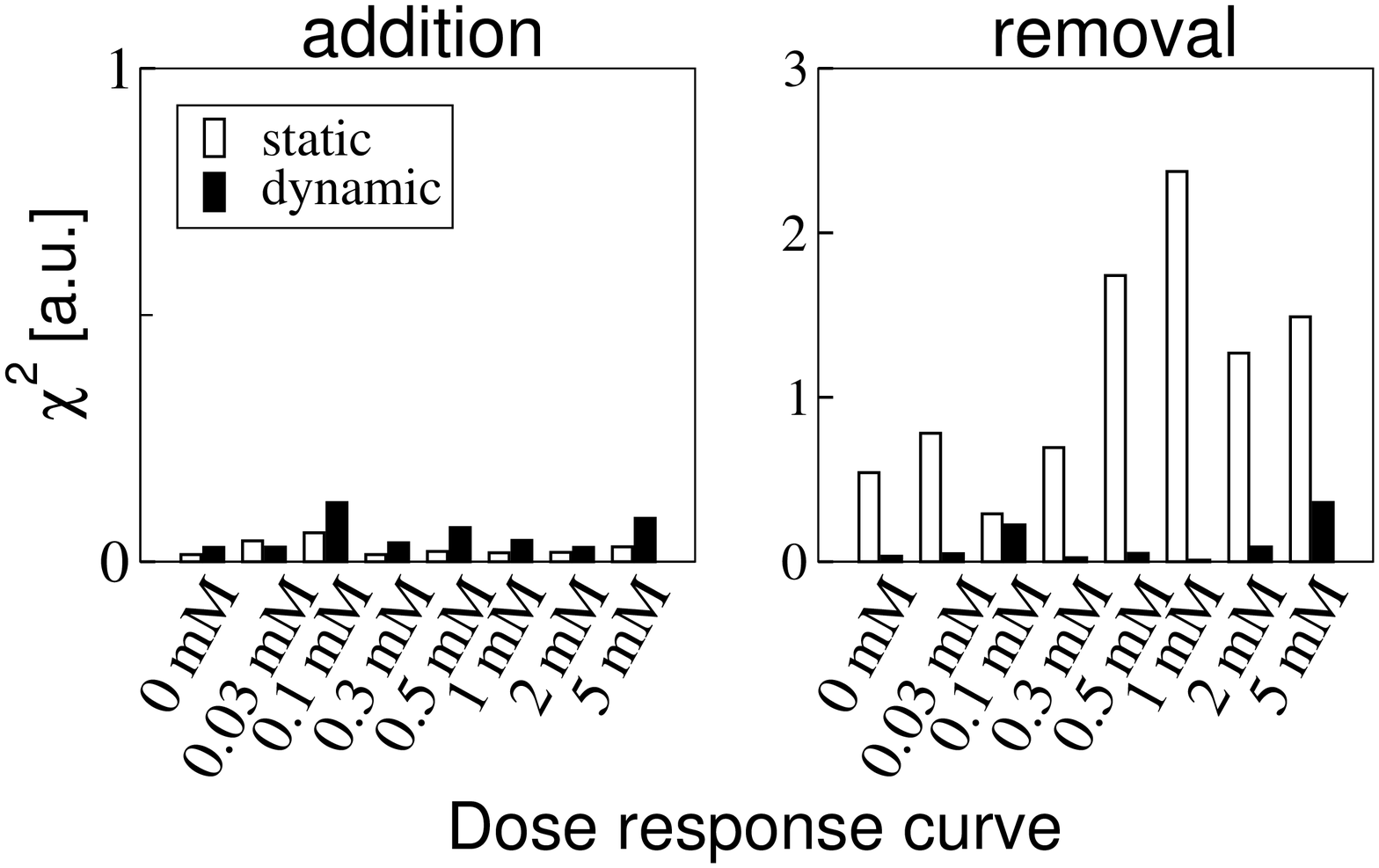}
\caption{\label{fig:residual_errors} Residual absolute squared errors per addition (left panel) and removal (right panel) dose-response curve for the static and dynamic MWC model as shown in Fig.~1 in the main text. Note the different axis scales for addition and removal plots. 
}
\end{figure}
\begin{figure}[t]
  \centering
    \includegraphics[width=0.5\textwidth]{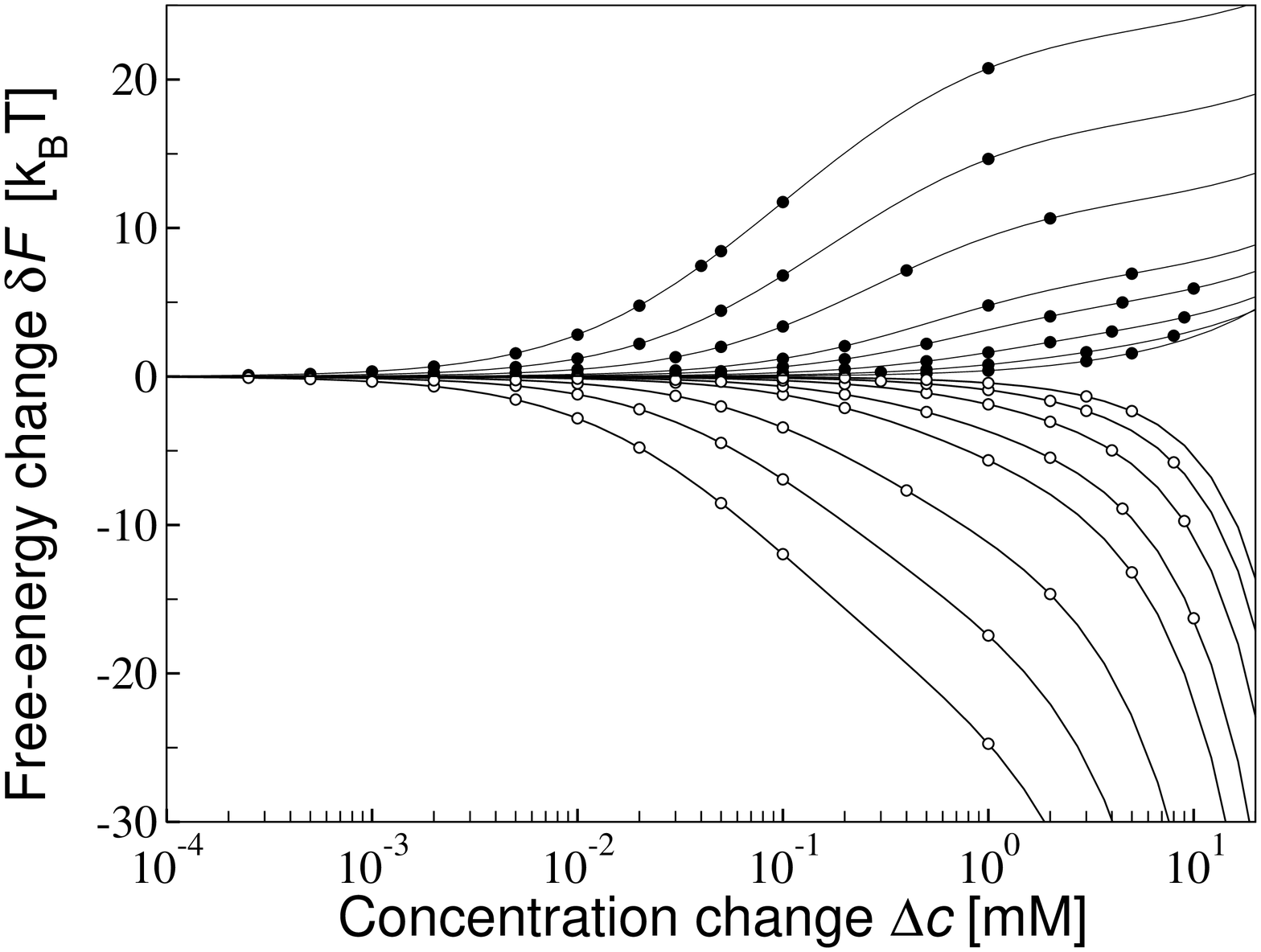}
\caption{\label{fig:dF}Changes in the free-energy difference $\delta F=F-F^*$ of a mixed-receptor complex upon concentration step changes $\Delta c$ of MeAsp (lines), where $F^*$ is the adapted free-energy difference. The curves correspond to ambient concentrations $c_0$=0, 0.03, 0.1, 0.3, 0.5, 1, 2 and 5 mM with free-energy differences for experimental concentration step changes indicated by symbols (filled circles for addition, open circles for removal of MeAsp).
}
\end{figure}

Figure~\ref{fig:residual_errors} quantifies the difference between measured dose-response curves and the static, as well as the dynamic MWC model, respectively, as detailed in the main text (cf. Fig.~1 in main text). We plot squared errors for each addition and removal dose-response curve. While the error for the dynamic MWC model is slightly larger for addition curves, its error for removal curves is much smaller than that for the static MWC model. Hence, the dynamic MWC model is suited better to describe the experimental data.

Figure~\ref{fig:dF} shows the free-energy change associated with each concentration step change. For increasing ambient concentrations, the free-energy changes generally decrease at a fixed concentration step change $\Delta c$. This is the reason for the reduced response amplitudes in the dynamic MWC model at large MeAsp step removals, because adaptation compensates for smaller free-energy changes at increasing ambient concentrations.

\section{Parameters for static and dynamic MWC model}

In Tab.~\ref{tab:figmaintext_parameters}, we list all parameters of the static and dynamic MWC model used for Fig.~1-3 in the main text and Fig.~\ref{fig:DR_bestfit}A in the Supplementary Text S1, as well as Fig.~\ref{fig:DR_bestfit}B and \ref{fig:alternativereceptormodels}A in Text S1.
\begin{table}[t]

\caption{\label{tab:figmaintext_parameters} Fitting parameters for the static and dynamic MWC model. Parameters include dissociation constants for Tar and Tsr receptors in the {\it on} and {\it off} states, respectively, $K^{\text{off}}_a$, $K^{\text{on}}_a$, $K^{\text{off}}_s$, $K^{\text{on}}_s$~\cite{KeyEndSko06}, the parameters of the linear approximation of the dependence of the receptor complex size on ambient concentration, $N(c_0) = a_0 + a_1 c_0$, as well as methylation and demethylation constants, $g_R$ and $g_B$ in Eq.~2 in the main text, respectively. Fitted parameters are indicated by crosses.
}
\begin{center}
\begin{tabular}{|p{2cm}||p{1.8cm}|p{0.5cm}|p{1.8cm}|p{0.5cm}||p{1.8cm}|p{0.5cm}||p{1.8cm}|p{0.5cm}|}
\hline
\multirow{2}{*}{Parameter} & \multicolumn{4}{|c||}{Fig.~1-3 (main text), \ref{fig:DR_bestfit}A (Text S1) } & \multicolumn{2}{|c||}{Fig.~\ref{fig:DR_bestfit}B (Text S1)} & \multicolumn{2}{|c|}{Fig.~\ref{fig:alternativereceptormodels}A (Text S1)}\\
& \multicolumn{2}{|c|}{static} & \multicolumn{2}{|c||}{dynamic} & \multicolumn{2}{|c||}{dynamic (best fit)} & \multicolumn{2}{|c|}{static (best fit)} \\
\hline
$K^{\text{off}}_a$ [mM]		& 0.02 		&  	& 0.02 		&  	& 0.02	&  x	& 0.056	&  x 	\\
$K^{\text{on}}_a$ [mM]  	& 0.5 		&  	& 0.5 		&  	& 0.50	&  x	& 0.15	&  x 	\\
$K^{\text{off}}_s$ [mM] 	& 100		&  	& 100		&  	& 216	&  x	& 100	& 	\\
$K^{\text{on}}_s$ [mM]  	& 10$^6$	&  	& 10$^6$	&  	& 10$^6$&  x	& 10$^6$& 	\\
\hline
$ a_0$				& 17.5		&  x	& 17.5		&  	& 22	&  x	& 37	&  x 	\\
$ a_1$  [mM$^{-1}$]		& 3.35		&  x	& 3.35		&  	& 9.6	&  x	& -0.78	&  x 	\\
\hline
$g_R$ [s$^{-1}$]		& N/A		& 	& 0.0069	&  	& 0.0079& 	& N/A	& 	\\
$g_B$ [s$^{-1}$]		& N/A		& 	& 0.11		&  x	& 0.127	&  x	& N/A	& 	\\
\hline
\end{tabular}
\end{center}
\end{table}

\section{Effect of receptor complex size on data collapse}

\begin{figure}[t]
  \centering
    \includegraphics[width=0.5\textwidth]{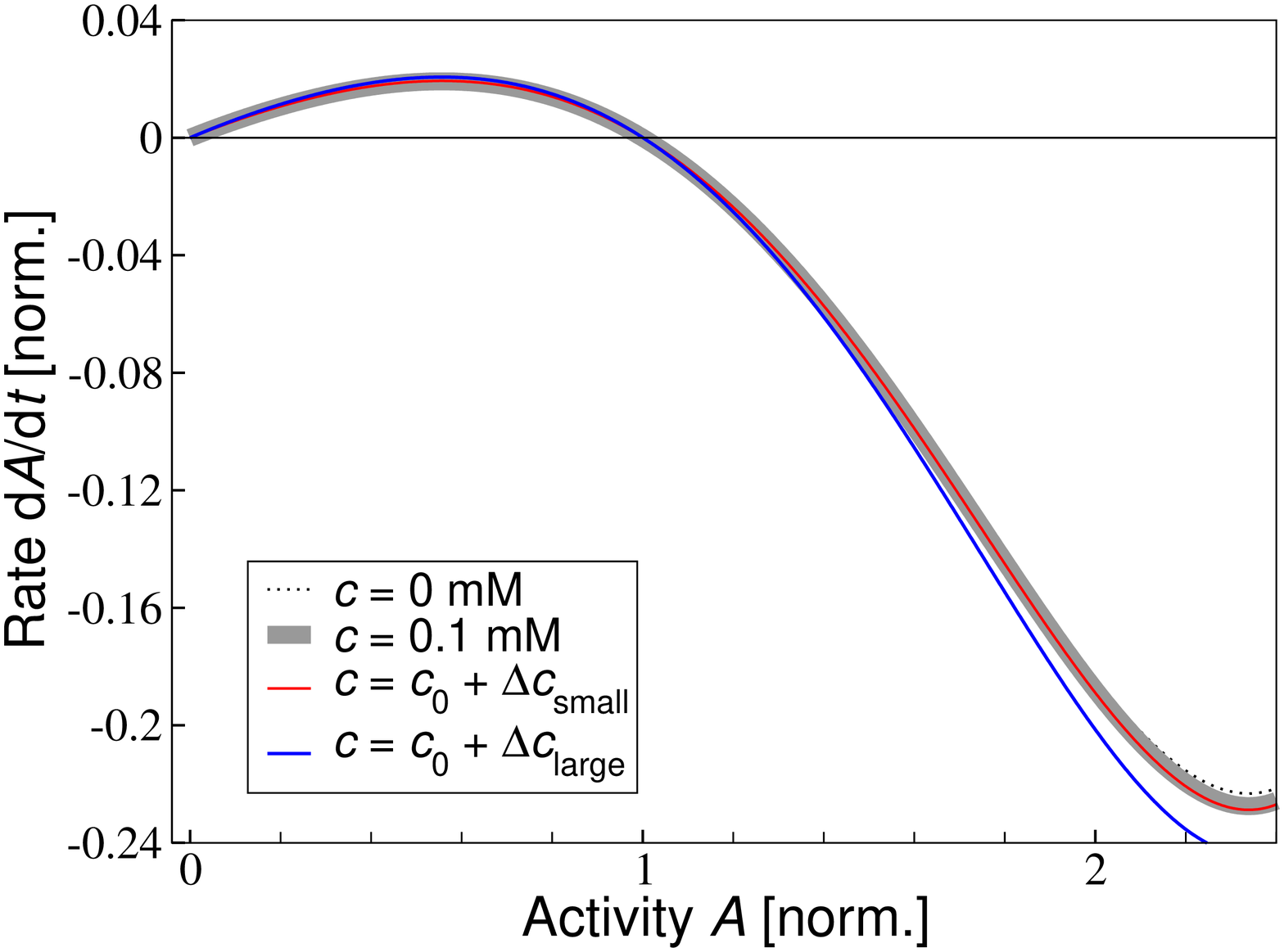}
\caption{\label{fig:N_datacollapse} Effect of ligand concentration-dependent receptor complex size $N(c)$ on the predicted data collapse according to Eq.~4 in the main text. Plotted are the predicted functions $f(A)$ for $N$ corresponding to ambient concentration $c=0.1$ mM (thick gray line), zero ambient (buffer; black dotted line), and concentration upon small (red line) and large (blue line) concentration changes as in Fig.~3 in the main text.} 
\end{figure}

We found from fitting the MWC model to dose-response curves from FRET that receptor complex size increases with ambient concentration (Fig.~2A in the main text). Hence, we would like to determine how the data collapse depends on this effect.
According to Eq.~4 in the main text, the rate of activity change is proportional to the receptor complex size $N$. As we do not have a model which describes how receptor complex size changes in time in response to concentration changes, we plot in Fig.~\ref{fig:N_datacollapse} the data collapse for different $N$ corresponding to the concentrations used in the experiments. This provides the envelope in which the data collapse is expected to change with $N$. We find that the data collapse does not change very much compared to the data collapse for ambient concentration $c_0$, and hence we neglected the effect of changing complex size in Fig.~3 in the main text.

\section{Unsuitable receptor signaling models}

We tested four alternative models for receptor signaling in an attempt to find a model, which describes the gradually reduced response amplitudes upon MeAsp removals at increasing ambient concentrations (cf. Fig.~1 in main text) without relying on adaptation and MeAsp dynamics. Dose-response curves for each model are shown in Fig.~\ref{fig:alternativereceptormodels}. We found none of the models produced a satisfying fit to the experimental data.

\subsection*{A Saturation model}
While ligand dissociation constants for the Tar receptor were previously determined from FRET data~\cite{KeyEndSko06}, slightly different values may lead to saturation of Tar receptors at smaller concentrations of MeAsp and reduced response amplitudes.
Figure~\ref{fig:alternativereceptormodels}A shows a fit of the static MWC model to addition as well as removal data. We fitted the parameters of the linear relationship between the receptor complex size and ambient concentration $c_0$, as well as the ligand dissociation constants for the Tar receptor, $K^{\text{on}}_a$ and $K^{\text{off}}_a$.
We find an unsatisfying fit, especially for the response to addition of MeAsp. Furthermore, the determined receptor complex size decreases
with ambient concentration (see {\it Inset}). This contradicts experiments which indicate an increasing receptor complex size~\cite{EndOleWin08}, as well as stabilization of polar receptor clusters with increasing receptor methylation level (corresponding to increasing ambient concentration)~\cite{ShiBanKaw05}.

\subsection*{B Imprecise adaptation model}

Figure~\ref{fig:alternativereceptormodels}B shows the effect of imprecise adaptation on the response amplitudes. For simplicity, we assume a linear decline of the adapted activity $A^*(c_0)$ with increasing ambient concentration $c_0$, with $A^*(0)=1/3$ and a 20 percent imprecision at concentration 10 mM (see {\it Inset}).
We observe that imprecise adaptation has only a small effect on the response amplitudes. Furthermore, imprecise adaptation tends to increase response amplitudes at high ambient concentrations due to normalization by a decreasing value of $A^*(c_0)$.

\subsection*{C Phase-separation model}

\begin{figure}[htp]
  \centering
    \includegraphics[]{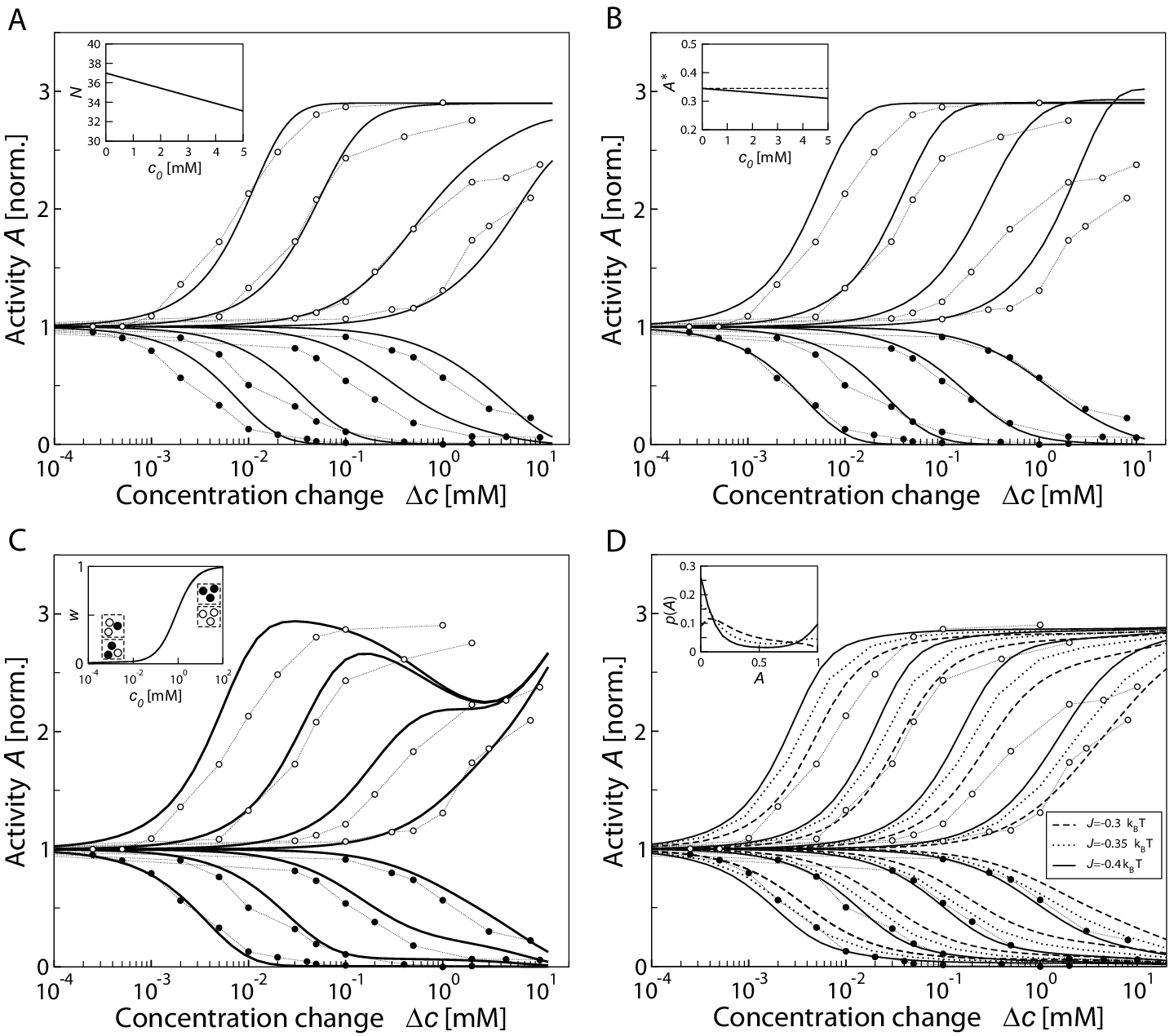}
\caption{\label{fig:alternativereceptormodels}Alternative models for receptor signaling.
(A) Saturation model. MWC model with optimized dissociation constants $K^{\text{off}}_a=0.056$ mM and $K^{\text{on}}_a=0.15$ mM. (A {\em Inset}) Total receptor complex size $N=a_0 + a_1 c_0$ for fitted parameters $a_0$=37 and $a_1$=-0.78 mM$^{-1}$.
(B) Imprecise adaptation model.
(B {\em Inset}) The steady-state activity decreases with ambient concentration, $A^*(c_0)/A^*(0) = 1 - (0.2 / 10\text{ mM})c_0$, where 20\% imprecision is reached at 10 mM. 
(C) Phase-separation model. Receptor complexes are found in separated complexes with probability $w$ depending on the ambient concentration $c_0$. Receptor complex size is assumed constant, $N=$18, for mixed and homogeneous receptor complexes.
(C~{\em Inset}) The probability $w(c_0)= p_1+p_2 \frac{c_0}{c_0+p_3}$, with $p_1$=0.01, $p_2$=0.99, $p_3$=0.8 mM.
(D) Receptor lattice model. Mixed trimers of Tar and Tsr dimers are arranged on a 4$\times$4 square lattice with periodic boundary conditions. The average activity of the lattice was calculated by exact enumeration. An attractive interaction between neighboring trimers in the same state was assumed, with interaction energy $J$=-0.4 $k_B T$ (solid line), $J$=-0.35 $k_B T$ (dotted line), and $J$=-0.3 $k_B T$ (dashed line).
(D {\em Inset}) Corresponding distributions of activities from all states (lattice configurations) when adapted to zero ambient concentration.
}
\end{figure}

In this model, a fraction $w$ of mixed receptor complexes composed of Tar and Tsr receptors form homogeneous receptor complexes of only Tar and only Tsr receptors at high ambient concentrations.
This separation reduces activity amplitudes at concentrations below the ligand dissociation constant $K_s^{\text{off}}$ for Tsr, as complexes of Tsr do not contribute to the response.
The total activity from mixed and homogeneous receptor complexes is
\begin{equation}
 A = \left[ 1-w(c_0) \right] A^{\text{mixed}} + w(c_0) \left( \nu_a A^{\text{Tar}} + \nu_s A^{\text{Tsr}} \right).
\end{equation}
The individual activities of mixed, $A^{\text{mixed}}$, and homogeneous receptor complexes of Tar, $A^{\text{Tar}}$, and Tsr, $A^{\text{Tsr}}$, were calculated according to the static MWC model. Mixed receptor complexes are composed of Tar and Tsr with ratio $\nu_a:\nu_s$=1:1.4. Homogeneous receptor complexes of Tar and Tsr, respectively, have the same ratio.
The resulting dose-response curves for this model are shown in Fig.~\ref{fig:alternativereceptormodels}C, assuming the probability of separation $w(c_0)$ from the {\it Inset}.
As the ambient concentration does not correspond nicely to the data points with decreasing response,  we did not find a well-fitting function $w(c_0)$.
Furthermore, this model predicts a smaller response to MeAsp when cells are pre-adapted to a ligand for which Tsr, but not Tar is sensitive (e.g. Serine).
This contradicts experiments, which show that cells remain sensitive (Ref.~\cite{SouBerg04} and V.S., manuscript in preparation).

\subsection*{D Receptor lattice model}

In the static MWC model the absolute cooperativity of the receptors in a complex results in a saturating response upon removal of attractant (cf. {\it Inset} of Fig.~1 in main text). Here, we consider an Ising lattice of $N_T$ two-state receptor trimers, where each trimer is coupled to neighboring trimers with finite interaction strength. We ask if a weaker coupling between receptors can describe the dose-response data, and in particular the reduced response amplitudes for removals. 

Figure~\ref{fig:alternativereceptormodels}D shows dose-response curves for different interaction strengths. We find, that in order to describe the addition data, a strong interaction between neighboring trimers has to be assumed. In this limit, the lattice model resembles the MWC model where all lattice sites are infinitely strongly coupled with all other receptors. In the {\it Inset} of Fig.~\ref{fig:alternativereceptormodels}D we show the activity distribution from all lattice states. As expected, the distribution becomes increasingly bimodal around the two states with {\it all receptors on} and {\it all receptors off}.

In the following we describe the details of the model and our simulations.
We used a 4-by-4 square lattice of mixed receptor trimers with periodic boundary conditions. 
Each trimer consisted of Tar and Tsr receptors with probabilities $\nu_{a}$ and $\nu_{s}$, respectively, where $\nu_{a}$:$\nu_{s}$=1:1.4 is the {\it in vivo} ratio of Tar and Tsr in a cell. The distribution of Tar and Tsr in trimers on the lattice was the same in all simulations.
Furthermore, each trimer has only two states, {\it on} and {\it off}. We numerate all possible states of the whole lattice (in total $n=2^{16}$ states for a 4-by-4 lattice, i.e. $N_T$=16 receptor trimers).
Assuming the lattice is in equilibrium, we can calculate the distribution of individual lattice states, and hence the average activity of the lattice. The probability of each lattice state depends on its energy, which has a contribution from the free-energy difference between the {\it on} and {\it off} states of each trimer and from the interaction between neighboring trimers.
The free-energy difference of trimer $j$ is computed according to the MWC model
\begin{equation}
 F^{j} = \epsilon (m^{j}) + \sum_{l=1}^{3} \ln \left( \frac{ 1+c/K^{\text{off}}_l}{ 1+c/K^{\text{on}}_l}\right),\label{eq:dF_trimer}
\end{equation}
where the index $l=a,s$ describes the receptor type, Tar or Tsr, within a trimer. The average methylation level of receptors in a trimer $j$ is denoted by $m^j$. The methylation energy is $\epsilon(m^j)=3\cdot (1-0.5\, m^j)$.

The interaction energy between neighboring trimers depends on their respective states. If they are in the same state (both {\it on} or both {\it off}), we assign the interaction energy $J$, if they are in different states, we assign the interaction energy $-J$. The total energy $E_k$ of a lattice state $k$ is determined by summing over all free-energy differences of individual trimers and interaction energies between neighboring trimers.

The methylation level $m^j$ of each trimer $j$ cannot be calculated analytically due to the finite coupling strength between receptor trimers, and hence was determined numerically using our adaptation model 
\begin{equation}
 \frac{d m^{j}}{dt} = g_R (1-A^{j}) - g_B (A^{j})^3.\label{eq:latticeadaptation}
\end{equation}
According to this model, the methylation level $m^{j}$ of the trimer $j$ depends on its average activity
\begin{equation}
 A^{j}= \frac{1}{Z} \sum_{k=1}^n s_k^{j} e^{-E_k}, 
\end{equation}
where $Z=\sum_k e^{-E_k}$ is the partition function, i.e. the sum over all lattice states and $s_k^j$ is the state (1={\it on}, 0={\it off}) of trimer $j$ in lattice state $k$.
The steady-state of Eq.~\ref{eq:latticeadaptation} determines the methylation level of each trimer, and therefore the adapted free-energy difference $\epsilon(m^j)$ in Eq.~\ref{eq:dF_trimer}.
The average activity of the whole lattice is determined by calculating the average trimer activity
\begin{equation}
 A=\frac{1}{N_T} \sum_{j=1}^{N_T} A^{j},
\end{equation}
where $N_T$ is the number of trimers on the lattice.

\section{Comparison of different adaptation models}

In Fig.~4 in the main text we compare different adaptation models to FRET data by collapsing the time courses, plotting the rate of activity change $dA/dt$ as a function of the activity $A$.
Here, we describe in detail the different models analyzed. In all of the models we assume precise adaptation, i.e. that methylation and demethylation rates only depend on the receptor complex activity.
For each adaptation model, we use a least-squares fit to the FRET data to determine the methylation and demethylation rate constants, assuming an adapted activity $A^*$=1/3 and receptor complex size $N=17.8$. The parameters and quality of fit $\chi^2$ for each of the models are listed in Tab.~\ref{tab:adaptmodels}.

Our model Eq.~2 in the main text
\begin{equation}
 \frac{dm}{dt}=g_R (1-A) - g_B A^3
\end{equation}
is denoted by ``$(1-A)$,~$A^3$'', referring to the activity dependence of the methylation and demethylation rates, respectively.
The best fit to the rate of activity change from FRET (fitting parameter $g_R$=0.0019 s$^{-1}$, resulting in $g_B$=0.030 s$^{-1}$ and quality of fit $\chi^2$=0.0021), and a representative time course for this model are shown in Fig.~4A and B in the main text, respectively (red solid lines). Note that this model describes the experimental data well, even at high activities. This model also shows a strong asymmetry in the time course with slow adaptation to addition and rapid adaptation to removal of MeAsp (cf. Fig.~2C in the main text). %check Ref 

We considered a variation of this model, denoted by ``$(1-A)$,~$A^2$'', without cooperativity of CheB-P molecules,
\begin{equation}
 \frac{dm}{dt} = g_R (1-A) - g_B A^2,
\end{equation}
where only one CheB-P molecule is necessary for demethylation of a receptor. Together with one factor $A$ from the activity of receptors, this leads to a demethylation rate proportional to $A^2$.
While this model is almost as well-suited to describe the rate of activity change from FRET as our main model~(fitting parameter $g_R$=0.0031 s$^{-1}$; $g_B$=0.017 s$^{-1}$, $\chi^2$=0.0022; see Fig.~4A in main text), the asymmetry of adaptation to addition and removal of MeAsp is less pronounced~(Fig.~4B in main text). Fitting dose-response data using this adaptation model resulted in adaptation rates which were much higher than observed in FRET time courses.
\begin{table}[t]
\begin{center}
\caption{\label{tab:adaptmodels} Parameters of the adaptation models when fitted to the rate of activity change from FRET shown in Fig.~4A in the main text. The size of receptor complexes was assumed to be $N=17.8$ in all models. $^a$ $K_1=K_r/[T]$ and $K_2=K_b/[T]$, where $K_r$=0.39 $\mu$M and $K_b$=0.54 $\mu$M are taken from Ref.~\cite{EmoClu08}. $^b$~$K_2 = K_b / [T]$ with $K_b$=1.25 $\mu$M~\cite{BarLei97}. The concentration of receptors is $[T]$=17~$\mu$M.}
 \begin{tabular}{|c||c|l||c|}
\hline
Adaptation model & $g_R$ (fitted) & other parameters &$\chi^2$\\ 
\hline
``$(1-A)$,~$A^3$''				& 0.0019 s$^{-1}$	& $g_B$=0.030 s$^{-1}$ & 0.0021 \\
\hline
``$(1-A)$,~$A^2$''				& 0.0031 s$^{-1}$	& $g_B$=0.017 s$^{-1}$ &  0.0022\\
\hline
``$\frac{1-A}{1-A+K_1}$, $\frac{A}{A+K_2}$''	& 0.0188 s$^{-1}$	& $g_B$=0.020 s$^{-1}$ & 0.0036	 \\
 &									& $K_1$= 0.0229$^a$	&	\\
 &									& $K_2$= 0.0318$^a$	& 	\\	
\hline
``$\frac{1-A}{1-A+K_1}$, $\frac{A^2}{A+K_2}$''	& 0.0046 s$^{-1}$	& $g_B$=0.014 s$^{-1}$ & 0.0025 \\
 &									& $K_1$= 0.0229$^a$	&	\\
 &									& $K_2$= 0.0318$^a$	& 	\\
\hline
``const, $\frac{A}{A+K_2}$''			& 0.00318 s$^{-1}$	& $g_B$=0.014 s$^{-1}$	& 0.0032\\
 & 									& $K_2$= 0.0735$^b$	& \\
\hline
\end{tabular}
\end{center}
\end{table}

Furthermore, the model denoted by ``$(1-A)$,~$A$'' without CheB-P feedback~\cite{EndWin06, HanCliWin08, VlaLovSou08, KalTuWu09}
\begin{equation}
 \frac{dm}{dt} = g_R (1-A) - g_B A
\end{equation}
yields the fitting parameter $g_R$=0.0048 s$^{-1}$, resulting in $g_B$=0.0091 s$^{-1}$ and quality of fit $\chi^2$=0.0025. Both, the fit of this model to the rate of activity change from FRET, and time courses, are described worse than with the other two models.

Another class of adaptation models was proposed in Ref.~\cite{EmoClu08} where the idea of ultrasensitivity to the adaptation dynamics of CheR and CheB-P was introduced. This model was proposed mainly for small changes in activity, such as for fluctuations around the steady-state activity due to noise.
We denote by ``$(1-A)/(1-A+K_1)$,~$A/(A+K_2)$'' the following model
\begin{equation}
 \frac{dm}{dt} = g_R \frac{1-A}{1-A+K_1} - g_B \frac{A}{A+K_2}.
\end{equation}
In this model, CheR (CheB) methylates (demethylates) inactive (active) receptors with Michaelis-Menten-type kinetics with Michaelis-Menten constant $K_1$ ($K_2$). In this model, there is no CheB-P feedback on the demethylation rate.
If $K_1$ and $K_2$ are small, the adaptation rate depends only weakly on the receptor activity. This results in long adaptation (relaxation) times, as well as strong sensitivity to protein fluctuations of either CheR or CheB through rates $g_R$ and $g_B$. We used $K_1=K_r/[T]=0.0229$ and $K_2=K_b/[T]=0.0318$, where we took $K_r$ and $K_b$ from~\cite{EmoClu08} and the concentration of receptors is $[T]$=17~$\mu$M.
As shown in Fig.~4A in the main text, the model without CheB-P feedback ``$(1-A)/(1-A+K_1)$,~$A/(A+K_2)$'' does not describe the rate of activity change from FRET (fitting parameter $g_R$=0.0188 s$^{-1}$; $g_B$=0.020 s$^{-1}$, $\chi^2$=0.0036). Furthermore, the time course shown in panel~B looks qualitatively different from experimental time courses (cf. Fig.~2C in main text).

A variant of the model also includes CheB-P feedback, which introduces another factor $A$ in the demethylation rate~\cite{EmoClu08}. We denote this model by ``$(1-A)/(1-A+K_1)$,~$A^2/(A+K_2)$'', which corresponds to
\begin{equation}
 \frac{dm}{dt} = g_R \frac{1-A}{1-A+K_1} - g_B \frac{A^2}{A+K_2}.
\end{equation}
This model fits the FRET activity change in Fig.~4A in the main text relatively well (fitting parameter $g_R$=0.0046~s$^{-1}$; $g_B$=0.014~s$^{-1}$, $\chi^2$=0.0025). However, this model is not very different from the simpler model ``$(1-A)$,~$A^2$'', as the CheB-P feedback introduces a strong activity-dependence.

% Barkai Leibler model
In the model suggested by Barkai and Leibler~\cite{BarLei97} CheR methylation does not depend on the activity state of receptors, and hence active, as well as inactive receptors get methylated. The kinetics of the methylation level is described by
\begin{equation}
 \frac{dm}{dt} = g_R - g_B \frac{A}{A+K_2},
\end{equation}
where the parameter value $K_2 = K_b / [T]$=0.074 with $K_b$=1.25 $\mu$M~\cite{BarLei97}, and $[T]$ as above.
Note that this model is a special case of above model ``$(1-A)/(1-A+K_1)$,~$A/(A+K_2)$'' with $K_1$=0.
Fitting to the FRET activity change yields $g_R$=0.00318 s$^{-1}$, resulting in $g_B$=0.014 s$^{-1}$ and quality of fit $\chi^2$=0.0032. The predicted data collapse, as well as time courses are very similar to the model ``$(1-A)/(1-A+K_1)$,~$A/(A+K_2)$'', and is therefore not plotted in Fig.~4 in the main text.

% ============================================================================
\section{Analysis of adaptation noise}

The receptor methylation level is subject to fluctuations due to the random nature of methylation and demethylation events. However, the adaptation dynamics also filters fluctuations in ligand concentration (translated into fluctuations of the receptor activity), averaging over and smoothing high-frequency noise by its slower dynamics.
Here, we estimate the variance of the methylation level of a receptor complex due to these two noise sources.
Equation~2 of the main text describes the deterministic kinetics of the average methylation level of receptors in a mixed receptor complex,
\begin{equation}
 \frac{dm}{dt} = g_R (1-A) - g_B A^3.\label{eq:dMdt_determ}
\end{equation}
Now, we consider the kinetics of the total methylation level of a receptor complex. The total methylation level $M$ is the sum of the individual methylation levels $m_i$ of all receptors in a complex, $M=\sum_{i=1}^N m_i$, with $N$ the number of receptors per complex. The rate of change of the total methylation level is
\begin{equation}
 \frac{dM}{dt} = N_R k_R (1-A) - N_B k_B A^3,\label{eq:dMdt}
\end{equation}
where we explicitly indicated the number of the modifying CheR and CheB-P molecules, $N_R$ and $N_B$, respectively. The modification rates for a single receptor are related to the rates for a receptor complex via $\tilde g_R=N_R k_R/N$ and $g_B=N_B k_B/N$, respectively.
To describe fluctuations about the mean total methylation level due to methylation and demethylation events, we introduce the noise $\eta(t)$ and write
\begin{equation}
 \frac{dM}{dt} = N_R k_R (1-A) - N_B k_B A^3 +  \eta(t).
\end{equation}
We assume $\eta(t)$ is the sum of individual noise terms contributed from each modifying enzyme CheR and CheB-P acting on groups of receptors, so-called {\it assistance neighborhoods}~\cite{LiHaz05,EndWin06,HanCliWin08},
\begin{equation}
\eta(t)= \sum_{i=1}^{N_R} \eta_{R(i)}(t) + \sum_{i=1}^{N_B} \eta_{B(i)}(t),
\end{equation}
where $\eta_{R(i)}$ and $\eta_{B(i)}$ are independent Gaussian white noises with zero mean $\langle \eta_{R(i)}(t)\rangle =\langle \eta_{B(i)}(t)\rangle$=0, autocorrelations $\langle \eta_{R(i)}(t) \eta_{R(i)}(t')\rangle = q_{R} \cdot \delta(t-t')$ and $\langle \eta_{B(i)}(t) \eta_{B(i)}(t')\rangle = q_{B} \cdot \delta(t-t')$, and vanishing cross-correlations.
To estimate the noise intensities $q_{R}$ and $q_{B}$, we assume that the number of methyl groups, which are added (removed) by each enzyme molecule CheR (CheB-P) in a time interval, are Poisson distributed, i.e. their variance equals the mean number of added (removed) methyl groups.
Therefore, the noise intensity $q_R$ associated with each CheR molecule is determined by its mean rate of methylation,
\begin{equation}
 q_{R} = k_R (1-A^*).
\end{equation}
Similarly, the noise intensity $q_B$ for demethylation is
\begin{equation}
 q_{B} = k_B {A^*}^3,
\end{equation}
where we only consider noise from one molecule of CheB-P.
We are interested in the steady-state fluctuations of the total methylation level. Therefore, we linearize Eq.~\ref{eq:dMdt} around the steady state to obtain the kinetics of the deviation $\delta M$ from the mean methylation level
\begin{eqnarray}
 \frac{d (\delta M)}{dt} & = & -\left(N_R k_R + 3 N_B k_B {A^*}^2\right)\delta A + \eta(t) \\
                         & = & -\left(N_R k_R + 3 N_B k_B {A^*}^2\right) \left( \frac{\partial A}{\partial F} \right) \left( \frac{\partial F}{\partial M} \cdot \delta M  + \frac{\partial F}{\partial c} \cdot \delta c \right) + \eta(t).
\end{eqnarray}
In the second step, we used that the receptor complex activity is subject to fluctuations from the methylation level, as well as the ligand concentration.
The derivative of receptor complex activity with respect to the free-energy difference (at steady state) is given by
\begin{equation}
 \frac{\partial A}{\partial F} = -A^* (1-A^*).
\end{equation}
The total methylation level of a receptor complex enters the free-energy difference through
\begin{equation}
 F = \underbrace{N -\frac{1}{2}M}_{=\sum_{i=1}^N \left(1-\frac{1}{2}m_i\right)} + \nu_a N \ln\left( \frac{1+c/K^{\text{off}}_a}{1+c/K^{\text{on}}_a}\right) + \nu_s N \ln\left( \frac{1+c/K^{\text{off}}_s}{1+c/K^{\text{on}}_s}\right),
\end{equation}
where $m_i$ are the methylation levels of receptors $i$.
Therefore, the derivative of the free-energy difference $F$ with respect to $M$ is given by
\begin{equation}
 \frac{\partial F}{\partial M} = -\frac{1}{2}.
\end{equation}
The derivative of the free-energy difference $F$ with respect to $c$ is given by
\begin{equation}
\frac{\partial F}{\partial c} = \nu_a N \left( \frac{1}{c+K^{\text{off}}_a} - \frac{1}{c+K^{\text{on}}_a}\right) + \nu_s N \left( \frac{1}{c+K^{\text{off}}_s} - \frac{1}{c+K^{\text{on}}_s}\right) \equiv \mu. 
\end{equation}
In summary, the kinetics of $\delta M$ is determined by
\begin{equation}
 \frac{d (\delta M)}{dt} = -\underbrace{\left(N_R k_R + 3 N_B k_B {A^*}^2 \right)A^* (1-A^*)}_{\equiv\lambda} \cdot \left( \frac{1}{2} \delta M - \mu \delta c \right) + \eta(t).\label{eq:deltam}
\end{equation}
To calculate the variance of the methylation level, we Fourier-transform Eq.~\ref{eq:deltam},
\begin{equation}
 i \omega \delta\hat{M} = -\lambda \left( \frac{1}{2} \delta\hat{M} - \mu \delta \hat{c} \right) + \hat{\eta},
\end{equation}
where the hat symbol denotes the Fourier transform.
The power spectrum $S_{M}$ of fluctuations in $M$ is defined as the average of the absolute value squared of $\delta \hat{M}$
\begin{equation}
 S_{M}(\omega) = \langle \vert\delta \hat{M}\vert^2 \rangle = \frac{ q_M + \lambda^2 \mu^2 \langle \vert \delta c \vert^2 \rangle }{\omega^2 + \lambda^2/4}.
\end{equation}
Here, $q_M$ denotes the noise intensity of methylation and demethylation, and $\lambda^2 \mu^2 \langle \vert \delta c \vert^2 \rangle$ is due to the uncertainty from the ligand concentration\footnote{Fluctuations of the ligand concentration characterized by $\langle \delta c^2 \rangle$ can be quantified as presented in Ref.~\cite{BergPurc77,BiaSet05} by
\begin{equation}
 \langle \delta c^2 \rangle = \frac{\alpha}{\pi a D \tau} \cdot c,\label{eq:ligandnoise}
\end{equation}
which corresponds to the time-averaged low-frequency limit of the noise power spectrum~\cite{BiaSet05,BiaSet08}. The parameter $a$ is the size of the ligand binding site of a receptor, $D$ is the ligand diffusion constant, and $\tau$ is an averaging time due to slower downstream reactions.
The parameter $\alpha$ is of the order one and depends on further receptor details~\cite{BiaSet05,BiaSet08}.
Using $\alpha\approx 1$, $a$=1 nm, $D$=100 $\mu \text{m}^2/\text{s}$, a typical ligand concentration $c=\sqrt{K_a^{\text{off}}K_a^{\text{on}}} = 0.1 \text{ mM}$~\cite{VlaLovSou08}, and $\tau=1/k_A=0.1$ s corresponding to slow autophosphorylation of CheA, we obtain $\langle \delta c^2 \rangle = 5\cdot 10^{-6}\text{ mM}^2$. 
}, where we assumed the two contributions are independent. In this formula, we see explicitly the noise filtering of fluctuations in ligand concentration by the kinetics of the methylation level, given by the frequency-dependent factor.

In the following, we calculate the variance of the methylation level of a receptor complex only due to methylation and demethylation events.
As $\eta(t)$ is composed of independent white noises, its total noise intensity $q_M$ is the sum of the individual noise intensities,
\begin{equation}
q_M=\langle \vert \hat{\eta} \vert^2 \rangle = {N_R} q_{R}+{N_B} q_{B} = 2 {N_R} q_{R} = 2 N_R k_R (1-A^*). 
\end{equation}
The last equality uses the fact that at steady state methylation and demethylation rates balance each other in Eq.~\ref{eq:dMdt}.
To calculate the variance of the methylation level we need to integrate the power spectrum over all frequencies $\omega$,
\begin{equation}
 \langle \delta M^2 \rangle = \int \frac{d \omega}{2\pi} \frac{ q_M }{\omega^2 + \lambda^2/4} = \frac{2 q_M }{\lambda},
\end{equation}
and obtain
\begin{eqnarray}
 \langle \delta M^2 \rangle & = & \frac{2 {N_R} q_{R}}{\left(N_R k_R + 3 N_B k_B {A^*}^2 \right)A^* (1-A^*)} = \frac{2 g_R}{\left(g_R + 3 g_B {A^*}^2 \right)A^* } = \frac{2}{A^* + 3 (1-A^*)}\label{eq:var_meth}\\
 & = & 0.87.\nonumber
\end{eqnarray}
Here, we used that the adapted activity is $A^*\approx 1/3$, and that the relation between the methylation and demethylation rate constants $g_R$ and $g_B$ is given by the steady state of the methylation kinetics Eq.~\ref{eq:dMdt_determ},
\begin{equation}
 g_B=g_R \frac{1-A^*}{{A^*}^3}.
\end{equation}

This result can be compared to results for other adaptation models previously reported in the literature.
Reference~\cite{HanCliWin08} uses a linear dependence of methylation and demethylation rates on the receptor activity, instead of the nonlinear dependence in Eq.~\ref{eq:dMdt_determ},
\begin{equation}
\frac{dm}{dt} = g_R (1-A) - g_B A.\label{eq:dMdt_determ_linear}
\end{equation}
In an equivalent approach using assistance neighborhoods as described above, the authors calculate the variance of the total methylation level to be
\begin{equation}
\langle \delta M^2 \rangle = \frac{1}{\vert \partial F / \partial M \vert} = 2. 
\end{equation}
Hence, the variance of the total methylation level of a receptor complex is reduced for adaptation kinetics with strong activity dependence of the demethylation rate (Eq.~\ref{eq:dMdt_determ}), compared to the linear adaptation model (Eq.~\ref{eq:dMdt_determ_linear}). The reason for this is the stronger negative feedback, leading to the rapid attenuation of fluctuations in the receptor complex activity. Mathematically, the prefactor of the linearized demethylation rate in Eq.~\ref{eq:var_meth} leads to the reduction of the variance of the methylation level of the receptor complex.

% ============================================================================
\section{Quantification of adaptation imprecision}

\begin{figure}[t]
  \centering
    \includegraphics[width=0.5\textwidth]{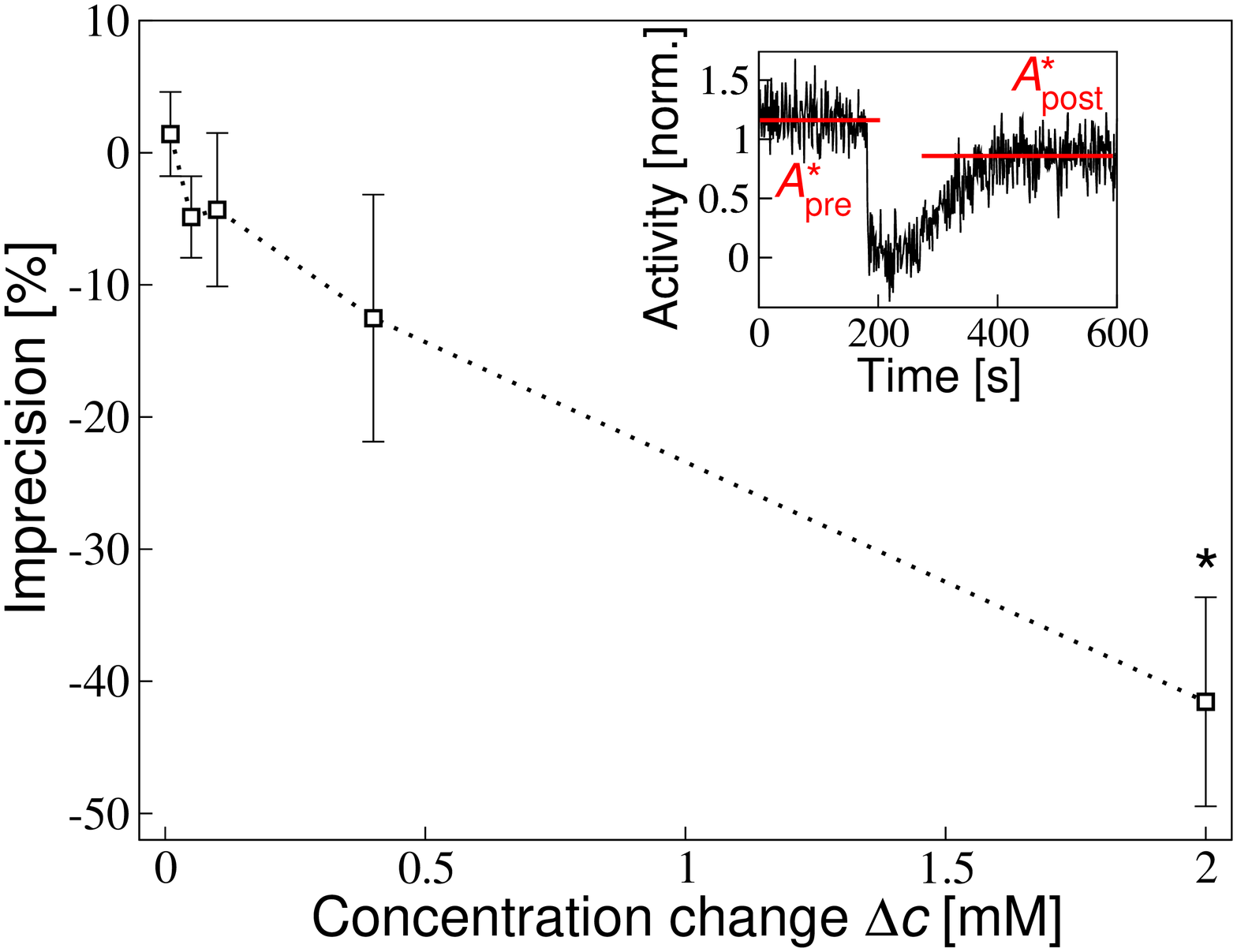}
\caption{\label{fig:imprecision} Imprecision of adaptation. FRET time courses were measured for cells adapted to 0.1 mM ambient concentration, and subject to various concentration step changes $\Delta c$. Levels of adapted FRET activity were determined before and after each added concentration step change, and the imprecision was calculated as $ (A^*_\text{post} - A^*_\text{pre}) /A^*_\text{pre}$. Symbols correspond to mean values of imprecision, and error bars indicate the standard mean error based on three replicates. The star indicates statistically significant difference from zero with Student's t-test p-value smaller than 0.05. ({\it Inset}) Example FRET time course for $\Delta c$=2~mM with adapted pre- and post-stimulus activity indicated. 
}
\end{figure}

In Fig.~\ref{fig:imprecision} we quantify the imprecision of adaptation. Cells were adapted to 100 $\mu$M ambient concentration with adapted pre-stimulus activity $A^*_\text{pre}$ measured by FRET. Concentration step changes of various sizes were added, and cells adapted to the new concentration with post-stimulus adapted activity $A^*_\text{post}$. We define a measure of imprecision as
\begin{equation}
 \text{Imprecision} = \frac{A^*_\text{post} - A^*_\text{pre}}{A^*_\text{pre}}.
\end{equation}
We find that adaptation is highly variable from experiment to experiment (high standard deviation). However, cells are found to  consistently adapt imprecisely at high concentrations.

% ============================================================================